\documentclass[superscriptaddress,prd]{revtex4} 
\usepackage{epsfig}
\usepackage{bm}
\usepackage{mathrsfs}
\usepackage{amsmath}
\usepackage{amsfonts}
\usepackage{amssymb}
\usepackage[T1]{fontenc} 
\usepackage{lmodern,aecompl}
\usepackage[latin1]{inputenc}
\usepackage{latexsym}
\usepackage{verbatim}
\usepackage{subfigure}
\usepackage{pstricks,pst-node,pst-text,pst-3d}
\def\bosy{\boldsymbol}

\begin{document}

\title{Chirikov and Nekhoroshev diffusion estimates: bridging 
the two sides of the river}

\author{Pablo~M.~Cincotta\email{pmc@fcaglp.unlp.edu.ar}}
\affiliation{Grupo de Caos en Sistemas Hamiltonianos,
Facultad de Ciencias Astron\'omicas y
Geof\'{\i}sicas, Universidad Nacional de La Plata and
Instituto de Astrof\'{\i}sica de La Plata (CONICET),
La Plata, Argentina}
\author{Christos Efthymiopoulos\email{cefthim@academyofathens.gr}}
\affiliation{Research Center for Astronomy and Applied Mathematics,
Academy of Athens, Greece}
\author{Claudia~M.~Giordano\email{giordano@fcaglp.unlp.edu.ar}}
\affiliation{Grupo de Caos en Sistemas Hamiltonianos,
Facultad de Ciencias Astron\'omicas y
Geof\'{\i}sicas, Universidad Nacional de La Plata and
Instituto de Astrof\'{\i}sica de La Plata (CONICET),
La Plata, Argentina}
\author{Mart\'{\i}n~F.~Mestre \email{mmestre@fcaglp.unlp.edu.ar}}
\affiliation{Grupo de Caos en Sistemas Hamiltonianos,
Facultad de Ciencias Astron\'omicas y
Geof\'{\i}sicas, Universidad Nacional de La Plata and
Instituto de Astrof\'{\i}sica de La Plata (CONICET),
La Plata, Argentina}

\begin{abstract}
We present theoretical and numerical results pointing towards a strong 
connection between the estimates for the diffusion rate along simple 
resonances in multidimensional nonlinear Hamiltonian systems that can 
be obtained using the heuristic theory of Chirikov and a more formal 
one due to Nekhoroshev. We show that, despite a wide-spread impression, 
the two theories are complementary rather than antagonist. 
{Indeed, although Chirikov's 1979 review has thousands of citations,
almost all of them refer to topics such as the resonance overlap
criterion, fast diffusion, the Standard or Whisker Map, and not
to the constructive theory providing a formula to measure diffusion
along a single resonance. However, as will be demonstrated explicitly
below, Chirikov's formula provides values of the diffusion coefficient
which are quite well comparable to the numerically computed ones,
provided that it is implemented on the so-called {\it optimal normal
form} derived as in the analytic part of Nekhoroshev's theorem.
On the other hand, Chirikov's formula yields unrealistic values
of the diffusion coefficient, in particular for very small values
of the perturbation, when used in the original Hamiltonian instead
of the optimal normal form.} In the present paper, we take advantage
of this complementarity in order to obtain accurate theoretical 
predictions for the local value of the diffusion coefficient along a resonance 
in a specific 3DoF nearly integrable Hamiltonian system. Besides, we 
compute numerically the diffusion coefficient and a full comparison of
all estimates is made for ten values of the perturbation
parameter, showing a very satisfactory agreement.

{\bf keywords}: Nekhoroshev's and Chirikov's diffusion theories--
chaos---instabilities--dynamics--Arnold diffusion
\end{abstract}
\maketitle

{\small{NOTICE: this is the author\'s version of a work that was accepted for publication in Physica D. Changes resulting from the publishing process, such as peer review, editing, corrections, structural formatting, and other quality control mechanisms may not be reflected in this document. Changes may have been made to this work since it was submitted for publication.}}

\section[]{Introduction}\label{intro}
In the present paper we analyze theoretically, and study by a concrete
numerical example, the connection between two different types of estimates
regarding the speed of local diffusion along a resonance of 
a nearly-integrable multidimensional nonlinear Hamiltonian system. These
are i) estimates based on the theory of diffusion developed by B. Chirikov
~\cite{Ch79}, and ii) estimates based on normal forms and the construction of
the Nekhoroshev theorem (~\cite{N77}, ~\cite{BGG85}, ~\cite{L92},~\cite{P93}).

The theory of Chirikov ~\cite{Ch79} relates the value of the diffusion coefficient
$D$ along a resonance, in a local domain of size $\epsilon^{1/2}$ around a
simply resonant point $\bm{I}^r$ of the action space, with the magnitude,
wavevector, and frequencies of the so-called driving harmonic terms in
systems of the form
\begin{equation}\label{eq1}
H(\bm{I},\bm{\theta})=H_0(\bm{I})
+\epsilon V(\bm{I},\bm{\theta}),\qquad \bm{I}\in \mathrm{G}\subset\mathbb{R}^N,
\quad \bm{\theta}\in \mathbb{T}^N,\quad \epsilon\ll 1,
\end{equation}
where $(\bm{I},\bm{\theta})$ are N--dimensional action--angle variables, 
$H_0$ is the integrable Hamiltonian and the  perturbation has the form
\begin{equation}\label{eq2}
\epsilon V(\bm{I},\bm{\theta})=\epsilon
\sum_{\bm{m}}V_{\bm{m}}(\bm{I})
\cos(\bm{m}\cdot\bm{\theta}),\qquad V_{\bm{m}}:\mathrm{G}\to\mathbb{R},
\qquad \bm{m}\in\mathbb{Z}^N/\{0\}~~.
\end{equation}

The theory of Chirikov has been reviewed in the framework of applications 
to dynamical astronomy in ~\cite{C02}. Using notations and 
terminology relevant to the present paper, we may summarize its main 
points as follows:

1) We decompose the Hamiltonian (\ref{eq1}) as
\begin{equation}\label{hamopt}
H(\bm{I},\bm{\theta})=Z(\bm{I},\bm{\theta})+R(\bm{I},\bm{\theta})~~
\end{equation}
where:

i) The function $Z(\bm{I},\bm{\theta})$ contains only resonant terms
associated with the particular resonance of interest (called by Chirikov
the `guiding resonance'), and can be written as:
\begin{equation}\label{nfres}
Z(\bm{I},\bm{\theta})=Z_0(\bm{I}) + \epsilon V_G\cos(\bm{m}_G\cdot\bm{\theta})
+ \ldots
\end{equation}
where $V_G$ and $\bm{m}_G$ are the amplitude and harmonic vector of the
main resonant term. After a change of basis, one requires furthermore that
the function $Z_0(\bm{I})$ be at least quadratic in a so-called resonant
action variable $p_1$, conjugate to a resonant angle $\psi_1=\bm{m}_G\cdot\bm{\theta}$. 
Then, the dynamics in the variables $(p_1,\psi_1)$ is given essentially by the 
pendulum dynamics (see section~\ref{sec4}).

ii) The term $R(\bm{I},\bm{\theta})$ contains harmonics of non-zero
wavevectors $\bm{m}$ not parallel to $\bm{m}_G$. Denoting by ${\cal V}$
the set of all such vectors, one has
\begin{equation}\label{remres}
R(\bm{I},\bm{\theta})=\sum_{\bm{m}\in{\cal V}}
V_{\bm{m}}(\bm{I})\cos(\bosy{m}\cdot\bm{\theta})~~.
\end{equation}

2) Estimates on the diffusion coefficient stem from examining how the
various terms in (\ref{remres}) affect the time evolution of certain
quantities, which represent exact integrals of motion of the Hamiltonian
flow under the term $Z(\bm{I},\bm{\theta})$, and approximate integrals
of the full Hamiltonian flow. The most important quantities of the
theory are: i) the energy of the pendulum part, and ii)
the remaining action variables.

In a first approximation, one computes the per-period (of the pendulum)
change of the values of the approximate integrals due to the various terms
in $R(\bm{I},\bm{\theta})$. In this evaluation, one approximates the time 
evolution of all angles in $R(\bm{I},\bm{\theta})$ by the ones corresponding 
to the evolution under the unperturbed separatrix solution of the pendulum.
One should consider phase correlations between
the various angles due to stickiness phenomena in the outer parts of the
separatrix-like chaotic layer. 
Integrating over the unperturbed separatrix 
solution for $\psi_1$ implies the use of Melnikov's integrals (see Appendix A).

3) Making the crucial assumption that the diffusion within the weakly
chaotic layers in the resonance web has a {\it normal} character,
the long-term variation of the approximate integrals can now be determined
in terms of the per-step variation of the same quantities. The final outcome
is a formula for the local value of the diffusion coefficient along the
resonance $\bm{m_G}$ in the vicinity of the point $\bm{I^r}$. After
some simplification, this formula reads

\begin{equation}\label{chi1}
D\lesssim  {1\over \Omega_G}\sum_{\bm{m}} \epsilon^2  |\omega_{\bm{m}}|\,
|V_{\bm{m}}(\bm{I}^r)|^2 e^{-\pi|\omega_m|\over \Omega_G}~~,
\end{equation}
where $\omega_{\bm{m}}\equiv \bm{m}\cdot\omega(\bm{I^r})$ and
$\Omega_G=(\epsilon V_G/|M_G|)^{1/2}$, with $M_G$ the nonlinear pendulum
mass, defined in section~\ref{sec4}. 
{The inequality in (\ref{chi1}) accounts for the fact that
the amplitudes $|V_{\bm{m}}|$ are not in the \emph{optimal form}, as will
be discussed in detail along this paper.
A variant of Chirikov's theory called the `stochastic pump' model 
was developed in ~\cite{1980AIPC...57..272T} and ~\cite{NYAS:NYAS119}. 

Regarding numerical implementations, in ~\cite{1980AIPC...57..323C} the authors  
computed the diffusion coefficient in a particular 2.5DoF nearly--integrable 
Hamiltonian system, whose unperturbed part is a bidimensional quartic oscillator. 
They obtained a good agreement between theory and experiment as long as the perturbation 
parameter was larger than a certain bound. However, the system considered depends on 
two  coupling parameters. A further example in the case of the 
so-called three body resonances in Solar System dynamics was provided in  
~\cite{CCFM10}. Agreement is again found beyond a certain 
bound in the perturbation.

Despite the large number of citations to Chirikov's 
report~\cite{Ch79}, by a systematic search we have been unable to identify 
other concrete quantitative applications of the same theory in the literature. 
In fact, most citations refer to the chaotic diffusion in the so-called 
resonance overlap regime, which occurs for sufficiently high values of 
$\epsilon$.

Nevertheless, the main goal of Chirikov's theory is to characterize 
the diffusion in the resonance web of multidimensional systems in the weakly 
chaotic limit, where there is no substantial resonance overlap. In this 
limit, the diffusion is conjectured to share features encountered in the 
mechanism of {\it Arnold diffusion}, proposed by Arnold ~\cite{A64}. 
However, Arnold's model is also a specific case with two parameters that 
can be varied independently one from the other. In contrast, in generic 
Hamiltonian systems the normal form theory introduces a dependence of all 
parameters that renders hardly tractable to generalize the proof of the 
existence of Arnold's mechanism ~\cite{L99}. In fact, although the diffusion 
in the web of resonances in the weakly chaotic regime has been observed in 
many numerical experiments, for instance~\cite{F70a}, ~\cite{F70b}, ~\cite{F72}, ~\cite{FS73}, 
~\cite{T82}, ~\cite{KK89}, ~\cite{WLL90}, ~\cite{L93}, ~\cite{DL93}, 
~\cite{SCP97}, ~\cite{ECV98}, ~\cite{LGF03}, ~\cite{GC04},  
~\cite{FGL05}, ~\cite{GLF05},~\cite{GLF06}, ~\cite{LGF09}, ~\cite{LGF10}, ~\cite{LGFb10},  
~\cite{CG12}, ~\cite{MCG_LAPIS11}, ~\cite{Mestre-PHD},  
~\cite{EH12}, not all these examples 
can be characterized as `Arnold diffusion'. In the sequel we consider systems 
satisfying the definition given in ~\cite{GLF05}, i.e. i) satisfying 
simultaneously the necessary conditions of the KAM and the Nekhoroshev theorems, 
and ii) being in the so-called `Nekhoroshev regime'. The first unambiguous 
numerical detection of local and global Arnold diffusion for such systems 
was made in ~\cite{LGF03} and ~\cite{GLF05}, respectively.

Our own main result presented below is the following: we will argue that,  
regarding the quantification of Arnold diffusion in such systems, Chirikov's 
and Nekhoroshev's theories meet and complement each other in an essential 
way, so that a proper implementation of Chirikov's theory requires computing 
first a so-called simply-resonant {\it normalized Hamiltonian function} which 
should be {\it optimal} in the Nekhoroshev sense. According to the Nekhoroshev 
theorem, the optimal normalized Hamiltonian function is computed via a recursive 
algorithm of canonical transformations, starting from the Hamiltonian (\ref{eq1}). 
Furthermore, this function has also the generic form of Eq.(\ref{hamopt}). 
However, the difference between the original and the optimal normalized 
Hamiltonian is that, in the latter case, all coefficients in the term 
$R(\bm{I},\bm{\theta})$ are bounded by a size exponentially small in an 
inverse power of $\epsilon$. Clearly, this affects also all the coefficients 
$V_{\bm{m}}(\bm{I})$ of the `driving' resonances, whose values appear in 
Chirikov's Eq.(\ref{chi1}). Working with a concrete numerical example, we 
then show that by using the optimal normalized Hamiltonian instead of the 
original one in Chirikov's formula, we can obtain precise estimates of the 
diffusion coefficient in the weakly chaotic limit. In fact, we compute 
such estimates and show their very satisfactory agreement with the values of 
the diffusion coefficient (for several values of $\epsilon$) found by a purely 
numerical integration of ensembles of orbits.

We note in this context that the connection between the Chirikov and Nekhoroshev 
theories is addressed by Chirikov himself in subsection 7.4 of ~\cite{Ch79}. 
In this review, Chirikov makes a qualitative discussion of how the optimal 
exponents appearing in the exponential estimates of Nekhoroshev theory affect 
the estimates of the speed of diffusion found in his own theory. This is further 
substantiated in subsection 7.6 of~\cite{Ch79}, by an analysis of the effects of 
higher order resonant terms on the diffusion rate in the weak perturbation limit. 
Here, instead, we provide direct evidence, that our normal form computation has 
reached an optimal order, and we also determine directly the effects of every 
driving resonance in the optimal Hamiltonian function using the exact version 
of Eq.(\ref{chi1}). 

The main steps of our study are as follows: we first perform an optimal 
simply-resonant normal form computation using a computer algebraic program, 
in order to study the diffusion in the thin chaotic layer in a domain of size 
$\epsilon^{1/2}$ along a particular simple resonance chosen by 
fixing the values of the action variables $\bm{I^r}\equiv(I_1^r,I_2^r,I_3^r)$ 
in the so-called  `perturbed 3DoF quartic oscillator model':
\begin{equation}\label{3Dpertquartic}
  \tilde{H}(\bm{y},\bm{x}) =
  \tilde{H}_0(\bm{y},\bm{x})+\epsilon \tilde{V}(\bm{x}),
\end{equation}
with
$$
\tilde{H}_0(\bm{y},\bm{x})
=\frac{1}{2}(y_1^2+y_2^2+y_3^2)+\frac{1}{4}(x_1^4+x_2^4+x_3^4)
,~~~\tilde{V}(\bm{x})=x_1^2(x_2+x_3)~~.
$$
We express (\ref{3Dpertquartic}) in action angle variables $(\bm{I},\bm{\theta})$ 
via a transformation $(\bm{y},\bm{x})\rightarrow (\bm{I},\bm{\theta})$ 
described in section~\ref{sec2}. Besides our acquaintance with its properties,  
(see~\cite{Mestre-PHD} and~\cite{MCG_LAPIS11}), our choice of model is motivated 
by our aim to compare results found here with those found in ~\cite{E08}, 
in which a different model was used ~\cite{FGL05}. Since extended studies of the 
diffusion in the weakly chaotic limit are available in this latter model as well 
(e.g.~\cite{GLF05} and ~\cite{LGF03}), we obtain in this way some indications 
regarding how general our results are. 

As explained above, the `bridge' between normal forms and the Chirikov approach 
is established after computing a simply resonant optimal normalized Hamiltonian 
valid in a neighborhood of a simply-resonant point $\bm{I^r}$ in the action 
space of the model (\ref{3Dpertquartic}). We then compute numerically the 
contributions of all the Melnikov integrals of the driving resonant terms, 
which we identify as the terms (except for one, see section~\ref{sec3}) 
appearing in the the so-called remainder function of the optimal normalized 
Hamiltonian. In other words, we identify the function $R(\bm{I},\bm{\theta})$ 
in Eq.(\ref{hamopt}) with the remainder function. Summing the values of the 
associated Melnikov integrals over all driving resonances we then arrive 
at a theoretical prediction for the value of the diffusion coefficient 
along the guiding resonance. We denote this value by $D_C$ (C stands for 
`Chirikov'), and we compute $D_C$ as a function of $\epsilon$ for ten  
values of $\epsilon$.

After computing $D_C(\epsilon)$ in the above way, we perform the following 
comparisons:

i) We compare $D_C(\epsilon)$ with the value of the diffusion coefficient
$D(\epsilon)$ computed by numerical experiments, i.e. by integrating ensembles 
of orbits with initial conditions in the thin chaotic layer surrounding the 
resonance in the neighborhood of $\bm{I^r}$. This particular calculation 
reveals one more salient feature of the normal form method: by transforming 
our orbital data into `good' action variables, obtained via a near-identity 
normalizing canonical transformation, we eliminate from the data all noisy 
behavior due to the so-called {\it deformation} effects (see e.g. ~\cite{G02}, 
p. 63). This removal proves to be a crucial step allowing to measure the 
diffusion due only to the {\it drift}, i.e. the slow motion along the 
resonance, after an integration time of the order of $t\leq 10^7$. Had 
we relied, instead, for this computation on the original action variables, 
whose time evolution reflects a combination of both the deformation and the 
drift effects, we would require much longer integration times (between $10^{9}$ 
and $10^{11}$ for the smaller values of $\epsilon$; see also ~\cite{LF02}). 
In addition, transforming the numerical data to good variables allows to 
reveal short-term features in the obtained diffusion curves which indicate 
up to what extent the diffusion can be considered as normal. In general, 
we find that the diffusion is indeed normal to a first approximation, but 
with secondary features representing possible deviations from the normal 
character. These, we aim to study in a future work. 

ii) As in ~\cite{E08} and ~\cite{EH12}, we check whether our data 
indicate a power-law relation between the diffusion coefficient and the 
normal form remainder $R$. In the present study we compare the size 
of the remainder $||R||$ with both $D$ and $D_C$. In both cases, we find 
a power-law of the form $D\sim ||R||^b$, or $D_C\sim ||R||^b$, with 
$b\approx 2.5$. This is somewhat smaller to the value 
$b\simeq 3$ found in ~\cite{E08} and in agreement with the results of  ~\cite{LGF10}.
 Regarding this latter point, we note that 
considering Chirikov's approach in combination with an optimal normal form 
construction instead of the original Hamiltonian (\ref{eq1}) leads to a 
simple argument of why the diffusion coefficient and the size of the 
remainder should be related by a power-law of the above form. If in 
Eq.(\ref{chi1}) we substitute the coefficients $V_m$ of the original 
Hamiltonian by the coefficients $f_m$ of the driving harmonics in the 
remainder of the optimal normalized Hamiltonian function, we arrive 
at the estimate 
\begin{equation}\label{chi2}
D_C\sim \sum_{\bm{m}} \epsilon^2 {|\omega_{\bm{m}}| 
\over \Omega_G} f_{\bm{m}}^2
e^{-\pi|\omega_{\bm{m}}|\over \Omega_G}~~.
\end{equation}
Here the sum is over wavevectors $\bm{m}$ labelling the driving 
resonances in the remainder function, while the relations 
$\omega_{\bm{m}}\equiv \bm{m}\cdot\omega(\bm{I^r})$ and $\Omega_G=(\epsilon 
V_G/|M_G|)^{1/2}$ still hold to the degree that the lowest order resonant term 
in the normal form and in the original Hamiltonian are practically the same. 
Now, from Nekhoroshev theory, for the largest $f_{\bm m}$ we have the estimate 
$f_{\bm m}\sim \exp[-(1/\epsilon^{1/2})^{1\over 1+\tau}]$, where $\tau$
is a positive constant (see \cite{E12} for a heuristic derivation of such 
estimates). The value of $\tau$ is determined by Diophantine bounds holding 
for the divisors $\bm{m\cdot\omega(I^r})$ appearing in all terms of the remainder. 
These bounds are of the form $|\bm{m\cdot\omega(I^r)}|>\gamma(I^{r})/|\bm{m}|^\tau$, 
where $\gamma$ is a positive constant and $|\bm{m}|$ is the $L_1$ modulus of 
$\bm{m}$. However, at the optimal normalization we also have an estimate for 
the minimum possible size of the {\it wavevectors} (the so-called `Fourier 
cut-off') of those terms in the remainder function containing the worst possible 
accumulation of divisors. This latter estimate reads 
$|\bm{m}|\approx\left(\epsilon^{1/2}\right)^{-1/(\tau+1)}$.
Combining these two estimates, along with the $O(\epsilon^{1/2})$ 
scaling of $\Omega_G$, and that $||R||\sim f_{\bm m}$ we get:
\begin{equation}\label{dcrem}
D_C\sim \epsilon^{\alpha} f_{\bm m}^2 e^{-(C_2/\epsilon^{1/2})^{1\over 1+\tau}}
\sim \epsilon^{\alpha} ||R||^{2+p}~~,
\end{equation}
with $\alpha$ an exponent that depends on $\tau$ and $p$ on the value 
of the constant $C_2$. Since $C_2\approx 1$, we also have $p\approx 1$ (see 
also ~\cite{EH12}). In ~\cite{E08}, the power-law relation 
between $D$ and $||R||$ was measured without any reference to Chirikov's theory. 
Thus, Chirikov's theory seems to provide a suitable framework for their 
interpretation. Further results regarding the connection between normal forms 
and estimates based on the Melnikov method (as are so the Chirikov estimates) 
can be found in ~\cite{N84}, ~\cite{MG97} and ~\cite{LGFc10}. 

The structure of the paper is as follows: in section~\ref{sec2} we briefly 
summarize the basic Hamiltonian model and choice of resonance in our 
study. In section~\ref{sec3} we discuss the normal form calculation. In 
section~\ref{sec4} we present the estimates on $D_C$ obtained via Chirikov's 
method. In section~\ref{sec5} we compare these results with numerical 
experiments, in which we show both the form of the diffusion curves found 
after transforming the numerical data to good normal form variables, 
as well as the main scalings, i.e. $D_C$ versus $D$, and $D_C$ versus
$R$. Section~\ref{sec6} is a summary of our main conclusions.

\section{Hamiltonian Model and choice of resonance}\label{sec2}

\subsection{Hamiltonian Model}\label{subsecA}
Our model consists of the Hamiltonian function (\ref{3Dpertquartic}) expressed 
in action--angle variables. To find the latter, we consider first the one-dimensional 
quartic oscillator
\begin{equation}
  \tilde{H}(y,x)=\frac{y^2}{2}+\frac{x^4}{4}~~.
\end{equation}
Let $h$ be the total energy and $a$ the associated oscillation amplitude, 
i.e. $h=a^4/4$. The solution $x(t)$ can be expressed in terms of the Jacobi elliptic 
cosine (${\rm cn}$) of modulus $k=1/\sqrt{2}$. Using the Fourier series development 
of the Jacobi elliptic cosine, we have~\cite{Ch79}:
\begin{equation*}
  x(t)=a \frac{\sqrt{2} \pi}{K_0}
  \sum_{n=1}^\infty
  \frac{1}{\cosh\big( (n-1/2)\pi \big)}
\cos\left( (2n-1)\frac{\pi a t}{2 K_0}\right),
\end{equation*}
where $K_0\equiv K(1/\sqrt{2})$ denotes the complete elliptic integral of the first 
kind. Introducing the following constants:
\begin{equation}
  \label{introd_constantes}
  \beta\equiv \frac{\pi}{2K_0}\approx 0.847213084793979,\qquad
  \alpha_n\equiv \frac{1}{\cosh\big( (n-1/2)\pi\big)}\qquad
  {\rm{and}}\quad \omega\equiv \beta a,
\end{equation}
we have
\begin{equation*}
  x(t)= 2^{3/2}\omega
  \sum_{n=1}^\infty
  \alpha_n \cos\big( (2n-1)\omega t \big).
\end{equation*}
The quantity $\omega$ is the fundamental frequency of the motion.
The coefficients $\alpha_n$ satisfy:
\begin{equation}\label{a23}
  \frac{\alpha_{n+1}}{\alpha_n} \approx \frac{1}{23}
  \qquad \mathrm{for}~n\geq1\qquad \rm{and}\qquad\alpha_1\approx 0.4.
\end{equation}
The relation between $\omega$ and $h$ is:
\begin{equation}
  \label{freq_energia_osc_cuart}
  \omega=\sqrt{2}\beta h^{1/4}.
\end{equation}
Since, in action--angle variables ($I,\bm{\theta}$) we have 
$\omega(I)=\frac{\partial H(I)}{\partial I}$, 
by means of Eq.~(\ref{freq_energia_osc_cuart}) we find 
\begin{equation*}
  h=A I^{4/3},\qquad {\rm{or ~equivalently}}\quad I=
\left(\frac{h}{A}\right)^{3/4},
\end{equation*}
where $A\equiv(3\beta/2\sqrt{2})^{4/3}\approx 0.867145326484821$. 
The dependence of the frequency on the action is given by:
\begin{equation}
  \label{frequency-action}
  \omega(I)=\frac{4}{3}AI^{1/3}~~.
\end{equation}
The cartesian coordinates can be finally expressed in terms of action--angle 
variables via the equations:
\begin{eqnarray}
  \label{osc_cuart_solucion_en_angulo-accion}
  x(I,\bm{\theta})=& (3\beta I)^{1/3}{\rm cn}
\left(\frac{\theta}{\beta},~\frac{1}{\sqrt{2}} \right)\nonumber \\ [3mm]
  y(I,\theta)=& \varrho \sqrt{2\left( AI^{4/3}
-\frac{1}{4} [x(I,\theta)]^4 \right)} ,
\end{eqnarray}
where  $\varrho$ stands for the sign of $y$ and its dependence on the 
angle is given by:
\begin{equation}
  \label{signo_p}
  \varrho\equiv
  \begin{cases}
   \quad 1  & \text{if } 0\leq \theta < \pi ,\\[2mm]
   -1  & \text{if } \pi \leq \theta < 2\pi.
  \end{cases}
\end{equation}
In the numerical computations, we also use the inverse transformation that allows 
to express the action--angle variables in terms of the cartesian variables
~\cite{MCG_LAPIS11}:
\begin{eqnarray}
  \label{cartesian_to_aa}
  I(y,x)~&=~& \left[\frac{1}{A} \left( \frac{1}{2}y^2
+\frac{1}{4}x^4 \right)\right]^{3/4},\\[3mm]
  \theta(y,x)&~=~&
  \begin{cases}
    \beta~{{\rm cn}^{-1}}\left(\frac{x}{[3\beta~I(y,x)]^{1/3}},
\frac{1}{\sqrt{2}} \right)
    & \text{if } y \geq 0, \\[2mm]
     2\pi - \beta~{{\rm cn}^{-1}} \left(\frac{x}{[3\beta~I(y,x)]^{1/3}},
\frac{1}{\sqrt{2}} \right)
    & \text{if } y < 0.
  \end{cases}
\end{eqnarray}

Passing now to the 3DoF Hamiltonian (\ref{3Dpertquartic}), by means of 
Eq.~(\ref{osc_cuart_solucion_en_angulo-accion}), we obtain a similar
equation for $x_j(I_j,\theta_j),\, j=1,2,3$ as the first in 
(\ref{osc_cuart_solucion_en_angulo-accion}). In a similar fashion 
Eq.~(\ref{frequency-action}) can be easily extended  to obtain the frequency 
vector in terms of the actions.
 
In action--angle variables the Hamiltonian is expressed as:
\begin{equation}
\label{3DHamiltoniano}
H(\bosy{I},\bosy{\theta}) = H_0(\bosy{I})+\epsilon V(\bosy{I},\bosy{\theta}),
\end{equation}
where
\begin{eqnarray}
  \label{3DHamiltoniano1}
  H_0(\bosy{I})&~=&~A ({I_1}^{4/3}+I_2^{4/3}+I_3^{4/3}),\\[2mm]
  V(\bosy{I},\bosy{\theta})
  &~=&~3\beta I_1^{2/3}{\rm cn}^2\left(\frac{\theta_1}{\beta},
~\frac{1}{\sqrt{2}} \right)
  \left[ I_2^{1/3} {\rm cn}\left(\frac{\theta_2}{\beta},
~\frac{1}{\sqrt{2}} \right)
    + I_3^{1/3} {\rm cn}\left(\frac{\theta_3}{\beta},
~\frac{1}{\sqrt{2}} \right) \right].\nonumber
\end{eqnarray}
This Hamiltonian system has been previously studied in 
~\cite{CGS03}, ~\cite{GC04} and  ~\cite{MCG09}. In the sequel we adopt 
a fixed value of the total energy:
\begin{equation}
   h\equiv 0.485\approx 1/{4\beta^4},
\end{equation}
which corresponds to a characteristic period (of the $x_2,x_3$ stable axial 
periodic orbits) very close to $2\pi$.

The perturbing potential can be developed in a Fourier series as:
\begin{equation}
  \begin{split}
    V(\bm{I},\bm{\theta})=&\hat{V}_{12}(\bm{I}) 
\sum_{n,m,k=1}^{\infty}\alpha_{nmk}
    \left\{\cos\big( 2(n+m-1)\theta_1\pm(2k-1)\theta_2 \big)
+\cos\big( 2(n-m)
      \theta_1\pm(2k-1)\theta_2 \big)\right\}+\\
    &+\hat{V}_{13}(\bm{I}) \sum_{n,m,k=1}^{\infty}\alpha_{nmk}
    \left\{\cos\big( 2(n+m-1)\theta_1\pm(2k-1)\theta_3 \big)
+\cos\big( 2(n-m)
      \theta_1\pm(2k-1)\theta_3 \big)\right\},
    \label{3DHamiltoniano1b}
  \end{split}
\end{equation}
with $\alpha_{nmk}\equiv \alpha_n \alpha_m \alpha_k\approx \alpha_1^3/23^{n+m+k-3}$ 
and $\hat{V}_{1j}(\bm{I})\equiv 2^{5/2}3\beta^4 I_1^{2/3} I_j^{1/3}$,
and  where the $\pm$ sign means that both terms are included in the series.

In ~\cite{MCG09} it is shown how to group all the coefficients $\alpha_{nmk}$ associated
to the same trigonometric function into a single coefficient $\alpha_{\bm{m}}$
that satisfies $\mathcal{O}(\alpha_1^3/23^2) \le \mathcal{O}(\alpha_{\bm{m}})\le 
\mathcal{O}(\alpha_1^3)$, in such a way that Eq.~(\ref{3DHamiltoniano1b}) can be 
rewritten as:
\begin{equation}
  V(\bm{I},\bm{\theta})=\hat{V}_{12}(\bm{I}) 
  \sum_{\bm{m}\in\mathcal{Y}}\alpha_{\bm{m}}
  \cos (\bm{m}\cdot\bm{\theta})+\hat{V}_{13}(\bm{I}) 
  \sum_{\bm{m}\in\mathcal{Z}}\alpha_{\bm{m}}
  \cos (\bm{m}\cdot\bm{\theta}) + \mathcal{O}(\alpha_1^3/ 23^3),
\end{equation}
where  $\mathcal{Y}$ and $\mathcal{Z}$ denote the subsets of wavevectors 
whose third and second
components, respectively, are zero. 

\subsection{Choice of resonance}\label{subsecB}

On applying the resonance condition, $\bm{m}\cdot\bm{\omega}(\bm{I})=0$, with
$\bm{m}\in \mathbb{Z}^3/\{\bm{0}\}$, to the unperturbed Hamiltonian, we get 
\begin{equation}
\label{3Dcondicionresonancia}
m_1 I_1^{1/3}+m_2 I_2^{1/3}+m_3 I_3^{1/3}=0~~.
\end{equation}
Thus, no resonant vector $\bm{m}$ can have all three components of the same sign. 
Besides the resonant vectors with two components equal to zero 
correspond to harmonics with null amplitude in the perturbation term of the 
original Hamiltonian. 

A point $\bm{I^r}$ in the action space is called a resonant point with respect 
to the resonance wavevector $\bm{m}$ if the three components $(I_1^r,I_2^r,I_3^r)$  
satisfy Eq.~(\ref{3Dcondicionresonancia}). By construction, every resonance 
wavevector is tangent, at $\bm{I^r}$, to the unperturbed energy surface, 
denoted as $\mathcal{I}_0$.

There are 12 resonant vectors of $\mathcal{O}(\epsilon)$ whose Fourier coefficient 
$\alpha_{\bm{m}}$, is at most of $\mathcal{O}(\alpha_1^3/23^2)$ and they are grouped 
in the following set:
\begin{equation*}
  \begin{split}
  \mathcal{V}_r(\epsilon,1/23^2)=\big\{&(2,-1,0),(2,-3,0),
  (2,-5,0),(4,-1,0),(4,-3,0),(6,-1,0),(2,0,-1),(2,0,-3),(2,0,-5),\\
&(4,0,-1),(4,0,-3),(6,0,-1)\big\}
  \end{split}
\end{equation*}
\begin{figure}[!ht]
  \centering
  \includegraphics[width=0.9\textwidth]{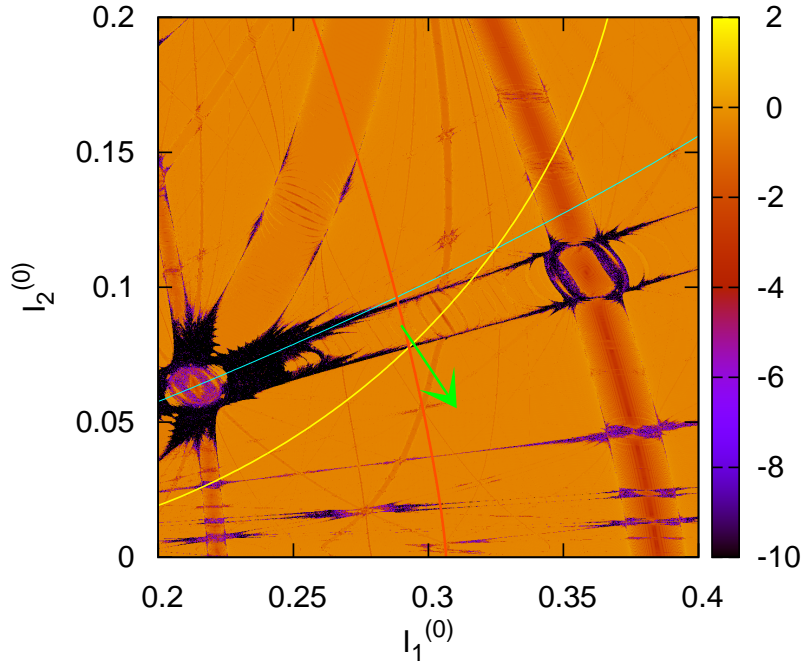}
  \caption{SALI map in action space, for $\epsilon=0.012$.
    The color palette is such that the more chaotic orbits
    appear in black while the regular ones in yellow.
    The green arrow starts at the point $\bm{I^r}$ and it  
    is parallel to the vector $(2,-3,0)$.}
  \label{guiding}
\end{figure}

The main resonances at order $\mathcal{O}(\epsilon^2,1/23^2)$ were obtained 
in ~\cite{MCG09}. Figure~\ref{guiding} shows a part of the resonant structure
as projected in the action plane $(I_1,I_2)$, for the given energy $h$, 
and $\epsilon=0.012$. The plot is obtained by computing the time evolution 
of $1000\times1000$ orbits with initial conditions $\theta_1=\theta_2=\theta_3=\pi/2$, 
and $I_1,I_2$ chosen in the domain $(I_1,I_2) \in [0.2,\,0.4]\times[0,\,0.2]$. 
The value of $I_3$ is obtained by solving the constant energy condition. 
In fact, after transforming to cartesian variables, we perform all numerical 
calculations in these variables, and back-transform, when needed, to action-angle 
variables. Along with the original equations, the variational 
equations of motion are integrated up to a time $t=10^4$. We compute then, for each orbit, the value 
of the so-called Smaller Alignment Index (SALI,~\cite{S01}), which is an indicator 
yielding the degree of regular or chaotic character of the associated orbit. 
That is,  the evolution in time of two different initial deviation vectors computing the norms of the difference d- (parallel alignment index) and the sum d+ (antiparallel alignment index) of the two vectors is followed, the time evolution of the smaller alignment index reflecting the chaotic or ordered nature of the orbit.
The use of chaotic indicators is known to yield an efficient method of depicting 
the resonant structure in the action space of multidimensional Hamiltonian
systems (see ~\cite{FL00} for the use of the FLI indicator, ~\cite{CS00} for 
the use of the MEGNO indicator, and ~\cite{LVE08} for the use of the APLE 
indicator in this framework). In the present case, using a color map of the 
values of the SALI indicator obtained for the various orbits in our grid, 
we obtain a clear representation of the web of resonances as well as the 
various chaotic layers appearing around each resonance.

In our study of diffusion, we focus on initial conditions (ICs) in the thin 
chaotic layer around the (guiding) resonance given by the wavevector:
\begin{equation}\label{mg}
\bm{m}_G\equiv(2,-3,0)~~.
\end{equation}
This resonance is projected close to the center of the plane $(I_1,I_2)$ of 
Fig.\ref{guiding}. Two more conspicuous and wide resonances cross transversally our 
resonance of interest. These are the resonances $(2,-1,-1)$, crossing $\bm{m}_G$ 
in the left part of the plot, and $(2,0,-2)$, crossing $\bm{m}_G$ in the 
right part of the plot. Many more resonances of smaller width, transverse 
to $\bm{m}_G$, are visible in the same plot. We have drawn
the center of the (unperturbed) high order resonances $(6,-7,-1)$ in cyan, 
$(6,-4,-3)$ in yellow and $(8,0,-7)$ in red. 
These resonances will play a significant role in Chirikov's formulation of the
diffusion along the guiding resonance.

In order to avoid as much 
as possible using initial conditions overlapping with the domains of
important resonance crossings, we chose to study the diffusion in the 
weakly chaotic layer surrounding the exactly resonant point 
$$
\bm{I}^r=(0.29, 0.08592592592592592, 0.434838361446344)
$$
of the action space. The associated resonant frequencies are
$$
\bm{\omega}^r \equiv \bm{\omega}(\bm{I}^r)=
(0.7652969051118440, 0.5101979367412294, 0.8759377456886241).~~
$$
We consider the following set of values for the perturbation parameter:
$$
\epsilon\in\mathcal{E}=
\{0.003, 0.005, 0.007, 0.008, 0.010, 0.012, 0.013, 0.015, 0.018, 0.020\}.~~
$$

In section~\ref{sec4} we present the results on estimates of the 
diffusion coefficient after an implementation of Chirikov's formula 
(\ref{chi2}) on the data obtained by a local simply-resonant normal 
form construction valid in a domain that contains the region
covered by our numerical orbits, around the above value of $\bm{I}^r$. 
Then, in section~\ref{sec5}, we compare these estimates with the results 
found by numerical integration of ensembles of orbits in the domain of 
interest.

\section{Normal form construction}\label{sec3}

In computing a resonant normal form for the dynamics in our domain of
interest, we used the same method as in ~\cite{E08}. The main
steps of the method are the following:

i) {\it Expansion around the center}: considering a union of polydisks
$|I_i'|<\rho$, where $\rho$ is a positive constant, and setting
$I_i'\equiv I_i-I_i^r$, where $\bm{I^r}$ is the central resonant value
in consideration, we perform an expansion of $H_0$ as a Taylor series
\begin{equation}\label{h0exp}
H_0=H_{0}^r+\bm{\omega}^r\cdot\bm{I}'
+ \sum_{i=1}^3\sum_{j=1}^3{1\over 2}M_{ij}^rI_i'I_j' +\ldots
\end{equation}
where $\bm{\omega}^r=\nabla_{\bm{I}} H_0(\bm{I}^r)$, while $M_{ij}^r$ are 
the entries of the Hessian matrix of $H_0$ at $\bm{I^r}$, denoted by $M^r$. 
Also, writing the perturbation as
\begin{equation}\label{h1four}
V(\bm{I},\bm{\theta})=\sum_{\bm{m}} h_{\bm{m}}(\bm{I})\exp(i\bm{m}\cdot\bm{\theta})
\end{equation}
in a domain where all three angles satisfy $0\leq Re(\theta_i)<2\pi$,
$|Im(\theta_i)|<\sigma$ for some positive constant $\sigma$, we expand all
the coefficients $h_{\bm{m}}$ around $\bm{I^r}$, namely
\begin{equation}\label{hkexp}
h_{\bm{m}}=h_{\bm{m}}^r+\nabla_{\bm{I}}h_{\bm{m}}^r\cdot \bm{I}'
+ {1\over 2}\sum_{i=1}^3\sum_{j=1}^3h_{\bm{m},ij}^rI_i'I_j' +\ldots
\end{equation}
Both series (\ref{hkexp}) and (\ref{h0exp}) have a common domain of
convergence around $\bm{I^r}$.

ii) {\it Action rescaling and book-keeping}:
We re-scale all action variables according to
\begin{equation}\label{resc}
J_i=\epsilon^{-1/2}(I_i-I_i^r)=\epsilon^{-1/2} I_i',~~~i=1,2,3
\end{equation}
so that all actions $J_i$ are $O(1)$ quantities in the domain
of interest. Since the transformation (\ref{resc}) is not
canonical, we multiply the Hamiltonian function by
$\epsilon^{-1/2}$ to restore correctness of the Hamiltonian
dynamics in the re-scaled action variables. Thus, the new
Hamiltonian reads:
${\cal H} (\bm{J},\bm{\theta})=\epsilon^{-1/2}H(\bm{I^r}
+\epsilon^{1/2}\bm{J},\bm{\theta})$.

We then split the Hamiltonian ${\cal H}(\bm{J},\bm{\theta})$ in terms of 
a similar order of smallness. In order to do so, we take into account 
the fact that the Fourier harmonics $\cos(\bm{m}\cdot\bm{\theta})$ in the 
Hamiltonian (\ref{3DHamiltoniano}) have amplitudes whose scaling is 
given essentially by Eq.(\ref{a23}). This implies an exponential decay 
factor $\sim e^{-\sigma|\bm{m}|}$ for a harmonic of order $|\bm{m}|$, 
where  $\sigma = 0.5\ln(23)$. Taking this fact into account we divide 
all harmonics in groups of a similar order of smallness, by introducing 
an integer constant 
\begin{equation}\label{kprime}
K'=\left[-{1\over 2\sigma}<\ln(\epsilon)>\right]~
\end{equation}
where $<\ln(\epsilon)>$ denotes the average value of $\ln(\epsilon)$
in the domain of values of $\epsilon$ considered in the present
study, namely from $\epsilon=0.003$ to $\epsilon=0.02$. In practice,
we take $K'=2$. Then, we re-write the Hamiltonian using a so-called
{\it book-keeping factor $\lambda$}, whose numerical value is $\lambda=1$,
as
\begin{eqnarray}\label{hamexp3}
{\cal H}(\bm{J},\bm{\theta})&=&\bm{\omega}^r\cdot \bm{J} + \lambda\epsilon^{1/2}
   \sum_{i=1}^3\sum_{j=1}^3{1\over 2}M_{ij}^rJ_iJ_j +\ldots
+\sum_{\bm{m}} \Bigg(\lambda^{1+[|\bm{m}|/K']}\epsilon^{1/2}h_{\bm{m}}^r\\
&+&\lambda^{2+[|\bm{m}|/K']}\epsilon\nabla_Ih_{\bm{m}}^r\cdot \bm{J}
+ \lambda^{3+[|\bm{m}|/K']}{\epsilon^{3/2}\over 2}
\sum_{i=1}^3\sum_{j=1}^3h_{\bm{m},ij}^rJ_iJ_j +\ldots\Bigg)
\exp(i\bm{m}\cdot\bm{\theta})\nonumber~~.
\end{eqnarray}\\
Setting $Z_0=\omega^r\cdot \bm{J}$, the Hamiltonian (\ref{hamexp3})
takes the form
\begin{eqnarray}\label{hamexpf}
{\cal H}(\bm{J},\bm{\theta})\equiv H^{(0)}(\bm{J},\bm{\theta})&=&Z_0+
\sum_{s=1}^{\infty}\lambda^s
H^{(0)}_s(\bm{J},\bm{\theta};\epsilon^{1/2})
\end{eqnarray}
where the superscript $(0)$ denotes, as usually, the original
Hamiltonian, and the functions $H^{(0)}_s$ are given by
\begin{equation}\label{h0s}
H^{(0)}_s = \sum_{\mu=1}^s\epsilon^{\mu/2}
\sum_{|\bm{m}|=K'(s-\mu)}^{K'(s-\mu+1)-1} H^{(0)}_{\mu,{\bm{m}}}(\bm{J})
\exp(i\bm{m}\cdot\bm{\theta})
\end{equation}
where $H^{(0)}_{\mu,{\bm{m}}}(\bm{J})$ are polynomials containing terms 
of degree $\mu-1$ or $\mu$ in the action variables $\bm{J}$. 
Precisely, we have:
$$
H^{(0)}_{\mu,\bm{m}}(\bm{J})=
\sum_{\mu_1=0}^{\mu-1}~~
\sum_{\mu_2=0}^{\mu-1-\mu_1}~~
\sum_{\mu_3=0}^{\mu-1-\mu_1-\mu_2}
{1\over\mu_1!\mu_2!\mu_3!}
{\partial^{\mu-1}h_{\bm{m}}(\bm{I^r})\over
\partial^{\mu_1}I_1\partial^{\mu_2}I_2\partial^{\mu_3}I_3}
J_1^{\mu_1}J_2^{\mu_2}J_3^{\mu_3}
$$
if $|\bm{m}|>0$, or
$$
H^{(0)}_{\mu,\bm{m}}(\bm{J})=
\sum_{\mu_1=0}^{\mu}~~
\sum_{\mu_2=0}^{\mu-\mu_1}~~
\sum_{\mu_3=0}^{\mu-\mu_1-\mu_2}
{1\over\mu_1!\mu_2!\mu_3!}
{\partial^{\mu}H_0(\bm{I^r})\over
\partial^{\mu_1}I_1\partial^{\mu_2}I_2\partial^{\mu_3}I_3}
J_1^{\mu_1}J_2^{\mu_2}J_3^{\mu_3}
$$
$$
~~~~~~~~+\sum_{\mu_1=0}^{\mu-1}~~
\sum_{\mu_2=0}^{\mu-1-\mu_1}~~
\sum_{\mu_3=0}^{\mu-1-\mu_1-\mu_2}
{1\over\mu_1!\mu_2!\mu_3!}
{\partial^{\mu-1}h_{0}(\bm{I^r})\over
\partial^{\mu_1}I_1\partial^{\mu_2}I_2\partial^{\mu_3}I_3}
J_1^{\mu_1}J_2^{\mu_2}J_3^{\mu_3}
$$
if $\bm{m}=0$.\\
\\

iii) {\it Resonant module:}
After choosing the vector of the guiding resonance 
according to Eq.~(\ref{mg}), we define the resonant module as the set
of all harmonics satisfying
\begin{equation}\label{resmod}
{\cal M}\equiv\left\{\bm{m}=0~\mbox{or}~\bm{m}//\bm{m}_G\right\}~~.
\end{equation}
The set ${\cal M}$ includes the wavevectors of all possible terms
appearing in the normal form.

{\it Hamiltonian normalization:} we perform Hamiltonian normalization
using a computer-algebraic program written by one of us (C.E.) in Fortran.
In this, we generate canonical transformations using the method of
Lie generating functions. For a review of the advantages
of this method from a computational point of view, see ~\cite{E12}. 

The normalization is performed in steps $r=1,2,...$, according to the
recursive formula
\begin{equation}\label{hr}
H^{(r)} = \exp(L_{\chi_r})H^{(r-1)}
\end{equation}
where $\chi_r$ is the r-th step generating function defined by
the homological equation
\begin{equation}\label{homo}
\{\omega^r\cdot \bm{J}^{(r)},\chi_r\}+\lambda^r
\tilde{H}^{(r-1)}_r(\bm{J}^{(r)},\bm{\theta}^{(r)})=0
\end{equation}
and $\tilde{H}^{(r-1)}_r(\bm{J}^{(r)},\bm{\theta}^{(r)})$ denotes 
all terms of $H^{(r-1)}$ which do not belong to the resonant module 
${\cal M}$. The operator $L_\chi$ is the Poisson bracket 
$L_\chi\equiv\{\cdot,\chi\}.$\\
\\
{\it Remainder and optimal normalization order:}
After $r$ normalization steps, the transformed Hamiltonian $H^{(r)}$
has the form
\begin{equation}\label{nfrem}
H^{(r)}(\bm{\theta},\bm{J})=Z^{(r)}(\bm{\theta},\bm{J};\lambda,\epsilon)
+R^{(r)}(\bm{\theta},\bm{J};\lambda,\epsilon)~~
\end{equation}
where $Z^{(r)}(\bm{J}^{(r)},\bm{\theta}^{(r)};\lambda,\epsilon)$ and
$R^{(r)}(\bm{J}^{(r)},\bm{\theta}^{(r)};\lambda,\epsilon)$ are the normal 
form and the remainder respectively. The normal form is a finite expression
which contains terms up to order $r$ in the book-keeping parameter
$\lambda$, while the remainder is a convergent series containing terms
of order $\lambda^{r+1}$ and beyond. Since in the computer we can only
store a finite number of remainder terms, we probe numerically the
convergence of the remainder function in the domain of interest in
the following way: writing the remainder in the form
\begin{equation}\label{rem1}
R^{(r)}(\bm{J}^{(r)},\bm{\theta}^{(r)}) = \sum_{s=r+1}^{\infty}
\lambda^s \sum_{|\bm{m}|} R_{\bm{m},s}^{(r)}(\bm{J}^{(r)})\exp(i\bm{m}\cdot\bm{\theta}^{(r)})
\end{equation}
we define the truncated norms
\begin{equation}\label{renorm}
||R^{(r)}(\xi)||_{\leq p}
= \sum_{s=r+1}^{p}
\sum_{|\bm{m}|}|{\cal R}^{(r)}_{\bm{m},s}(\xi)|
\end{equation}
where ${\cal R}^{(r)}_{\bm{m},s}(\xi) = R^{(r)}_{\bm{m},s}(J_1=J_2=J_3=\xi)$, and 
$\xi$ denotes a distance from the central point $\bm{I}^r$ in the 
action space, in re-scaled units, at which the diffusion is measured. 
By plotting the values of  $||R^{(r)}(\xi)||_{\leq p} $ versus $p$ we 
can have a numerical indication of whether the remainder function converges 
at $\xi$ after a certain value of $p$. Since the distance from the center 
to the chaotic layer is constant in the re-scaled action variables, by 
plots like in Fig.~\ref{guiding} we have estimated the value of $\xi=0.07$.

Figure {\ref{normal_form}}~(a) shows a calculation of this type 
for $\epsilon=0.01$. The middle curve, which corresponds to the normalization 
order $r=6$, shows the value of $||R^{(r)}(\xi)||_{\leq p}$ as a function of 
$p$ for $p=7,\dots,22$. Clearly, after $p=9$ the cumulative sum (\ref{renorm}) 
shows no further substantial variation, which indicates that the remainder 
series converges after just three consecutive terms $p=7,8$ and $9$.  
The lower  and upper curves show now the same effect for the normalization 
orders $r=11$ and $r=14$, respectively. The main effect to note is that the 
estimated remainder value $||R^{(r)}(\xi)||_{\leq 22}$ found for $r=14$ is 
larger than the one for $r=11$, implying that the {\it optimal} normalization 
order $r_{opt}$ is below $r=14$. Fig.{\ref{normal_form}}~(b) shows, 
precisely, the asymptotic character of the above normalization, showing 
$||R^{(r)}(\xi)||_{\leq 22}$ against the normalization order $r$ for various 
values of $\epsilon$ as indicated in the figure. We observe that in the 
considered range of values of $\epsilon$ the optimal normalization order 
turns to be always 10 or 11. In fact, it is known from basic theory that 
the optimal order of normalization is in general an increasing function 
of $1/\epsilon$. However, depending on the number--theoretical properties 
of the frequencies $\bm{\omega}^r$, the increase may occur by abrupt steps 
(see e.g. ~\cite{EGC04}). These steps are due to the fact that the smallest 
possible divisor $|\bm{m}\cdot\bm{\omega}^r|$ may remain invariant for long 
transient intervals of values of $|\bm{m}|$, before eventually being forced 
to follow the `envelope' provided by the Diophantine inequality  
$|\bm{m}\cdot\bm{\omega}^r|>\gamma/|\bm{m}|^\tau$. This phenomenon appears, 
precisely, for our present choice of resonant frequencies.  

\begin{figure}[!ht]
  \centering
  \includegraphics[width=0.6\textwidth]{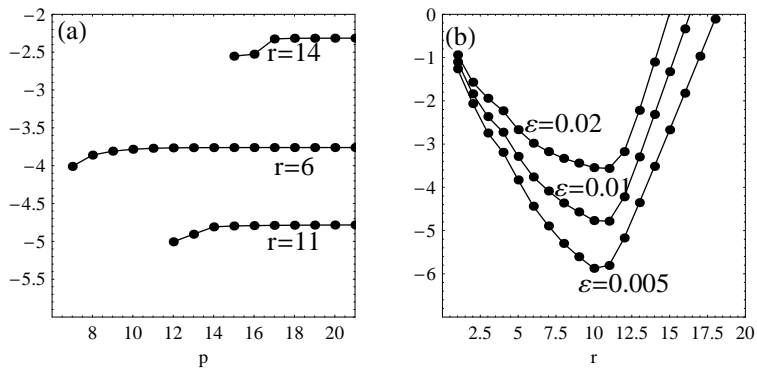}
  \caption{(a)~Values of $||R^{(r)}(\xi)||_{\leq p}$ as a function of $p$ for
three normalization orders, $r=6, 11, 14$. (b)~Values of the
remainder as a function of the normalization order $r$ for three values
of $\epsilon$ ~(right panel).}
  \label{normal_form}
\end{figure}

From the normal form calculation we retain three sets of data used
in subsequent calculations:

1) The value of the optimal remainder $||R^{(r_{opt})}||$ as a function of 
$\epsilon$ (found from the minima of all curves as in  
Fig.\ref{normal_form}~(b)). This is given in the Table I. 
\begin{table}[h!]
\centering
\begin{tabular}{cc}
\hline\hline
\vspace*{-2ex} \\ %
$\epsilon$\hspace{4mm} & $||R^{(r_{opt})}||$ \\ %
\hline %
 $0.020$\hspace{4mm} & $2.4\times 10^{-4}$\\%
\hline  %
 $0.018$\hspace{4mm} & $1.6\times10^{-4}$\\%
\hline  %
 $0.015$\hspace{4mm} & $7.7\times10^{-5}$\\%
\hline  %
 $0.013$\hspace{4mm} & $4.4\times10^{-5}$\\%
\hline  %
 $0.012$\hspace{4mm} & $3.3\times10^{-5}$\\%
\hline  %
 $0.010$\hspace{4mm} & $1.6\times10^{-5}$\\%
\hline  %
 $0.008$\hspace{4mm} & $7.1\times10^{-6}$\\%
\hline  %
 $0.007$\hspace{4mm} & $4.4\times10^{-6}$\\%
\hline  %
 $0.005$\hspace{4mm} & $1.3\times10^{-6}$\\%
\hline  %
 $0.003$\hspace{4mm} & $2.4\times10^{-7}$\\%
\hline\hline  \vspace*{-4ex} %
\end{tabular}
\caption{The value of the optimal remainder $||R^{(r_{opt})}||_{\leq 22}$ 
as a function of $\epsilon$ for all values $\epsilon\in\mathcal{E}$ according 
to the simply resonant normal form calculation performed as exposed above.}
\label{tabla3}
\end{table}

2) The form of the normalized Hamiltonian at the optimal normalization
order $r_{opt}=10$, including the remainder terms up to the order 22. 
This is transformed below in the basis used in Chirikov's theory in order
to compute the amplitudes and wavevectors of the guiding resonances.

3) The normalizing canonical transformation yielding the old canonical variables 
as functions of the new canonical variables. This transformation is provided 
directly by the composition of the computed Lie generating functions via the 
equations:
\begin{eqnarray}\label{newold}
q_{new}&=&\exp(-L_{\chi_1})\exp(-L_{\chi_2})
...\exp(-L_{\chi_r})q_{old}\nonumber\\
\end{eqnarray}
where $q_{old}$ and $q_{new}$ refer to anyone of the three action or angle 
variables before and after implementing the canonical transformations. In the 
case of the old action variables, we first compute the values of the re-scaled 
actions $\bm{J}=\epsilon^{-1/2}(\bm{I}-\bm{I^r})$ from the values of the 
original actions $\bm{I}$ which are available by our numerical data. Then, 
we `pass' the values of the actions $\bm{J}$ to the transformation 
(\ref{newold}).

\section{Diffusion estimates using Chirikov's theory}\label{sec4}

We will now use the data of the normal form computation exposed in the
previous section, in order to obtain estimates of the diffusion coefficient
using the theory of Chirikov. The reader is deferred to ~\cite{Ch79}
and~\cite{C02} for a detailed presentation of this theory. 

As a preliminary step, we re-express the optimal normalized Hamiltonian 
functions found in the above section, for each value of $\epsilon$ in 
the considered set $\mathcal{E}$, into a function expressed in non-scaled 
action variables $(I_j-I^r_j)=\epsilon^{1/2}J_j$, $j=1,2,3$. This is 
done by the back transform
\begin{equation}\label{nfremopt}
H^{(r_{opt})}(\bm{\theta},\bm{I-I^r})=\epsilon^{1/2}\Bigg[
Z^{(r_{opt})}(\bm{\theta},\epsilon^{-1/2}(\bm{I-I^r});\lambda,\epsilon)
+R^{(r_{opt})}(\bm{\theta},\epsilon^{-1/2}(\bm{I-I^r});\lambda,\epsilon)
\Bigg]~~. 
\end{equation}
The lowest order terms in the above expression are of the form
\begin{equation}
H^{(r_{opt})} = \bm{\omega_G}\cdot\bm{(I-I^r)} + O\left((\bm{I-I^r})^2\right) 
+...+\epsilon [V_G+O\left((\bm{I-I^r})\right) +...]
(\cos(\bm{m}_G\cdot\bm{\theta})+...
\end{equation}
The guiding resonance is the one given by $\bm{m}_G=(2,-3,0)$. For the 
constant $V_G$ we find the numerical value $V_G\simeq 0.005259$.

Following Chirikov's formulation, we  perform a `change of basis', i.e. 
define three new fundamental directions in the action space and re-write 
the action variables in terms of components in these new directions. 
This will be done through a canonical transformation and, to this end, 
we define the vectors  
\begin{equation*}
  \label{Chirikov_base}
  \begin{array}{c}
    \bosy{\mu}_1 = \bosy{m}_G,\qquad \bosy{\mu}_2 = 
\bosy{\omega}^r / |\bosy{\omega}^r|      \\[3.5mm]
    \bosy{\mu}_3 = ( \bosy{n}^r \wedge  \bosy{\omega}^r ) / 
|\bosy{n}^r \wedge   \bosy{\omega}^r |
=(0.6769019893644146, 0.2005635524042710,-0.7082216872148703),
\end{array}
\end{equation*}
where
$$
\bm{n}^r=
(\partial [\bm{m}_G\cdot\bm{\omega} (\bm{I})]/\partial\bm{I})_{\bm{I}^r}=
(1.759303230142170, -5.937648401729823,0),
$$
is a vector normal to the guiding resonance surface at the point $\bm{I}^r$.

Geometrically, we have that $\bm{\mu}_1$ lies in the tangent plane to
$\mathcal{I}_0$ at the point $\bm{I}^r$, $\bm{\mu}_2$ is normal to that plane 
and $\bm{\mu}_3$  is simultaneously orthogonal to $\bm{n}^r$ and to  $\bm{\mu}_2$, 
i.e. it is tangent, at $\bm{I}^r$, to the intersection between the guiding resonance
surface and $\mathcal{I}_0$. These three vectors are linearly independent if and 
only if $\bm{m}_G$ is not perpendicular to $\bm{n}^r$. A way to ensure this 
geometrical condition is to assume that $\mathcal{I}_0$ is convex at the point 
$\bm{I}^r$. This is easily checked to be true in our specific example. 

Let $\Upsilon$ $\in \mathbb{R}^{3\times3}$ be the matrix whose $i-$th row is 
the vector $\bosy{\mu}_i$, for $i=1,2,3$, and let $G$ be a generating function 
given by:
\begin{equation*}
  G(\bosy{p},\bosy{\theta})\equiv 
  \sum_{j=1}^3 
  \left( 
  I_j^r  
  + 
  \sum_{k=1}^3 p_k\Upsilon_{kj} 
  \right)
  \theta_j.
\end{equation*}
The associated canonical transformation $(\bm{I},\bm{\theta})\rightarrow 
(\bm{p},\bm{\psi})$ can be written explicitly as:
\begin{equation}
  \label{basetra}
  \begin{array}{l}
    \bosy{\psi} = \Upsilon \bosy{\theta},  \\[3.5mm]
    \bosy{p}    = \Upsilon^{-T}(\bosy{I}-\bosy{I}^r),
  \end{array}
\end{equation}
where $\Upsilon^{-T}\equiv (\Upsilon^T)^{-1}$ denotes the inverse matrix of 
$\Upsilon$'s transpose. This transformation consists of a translation of the origin 
to the resonant action, followed by a change of base in the action space.

The new momenta, $p_k$ ($k=1,2,3$), are the components of the vector 
$(\bosy{I}-\bosy{I}^r)$ in Chirikov's base such that, $p_1$ measures the deviation 
of the actual motion from  the resonant point {\it across} the guiding resonance 
layer, $p_2$ gives the unperturbed energy variation $H_0$, and $p_3$ measures the 
departure from the resonant value {\it along} the guiding resonance layer, i.e., 
in the direction along which we will measure the diffusion coefficient. 
The conjugate angle of $p_1$, $\psi_1= \bm{m}_G\cdot\bm{\theta}$, is the 
resonant angle.

After the transformation (\ref{basetra}), the truncated Hamiltonian (up to 
order 22) takes the form:

\begin{eqnarray}\label{hamppsi}
H(\bm{p},\bm{\psi})&=&{p_1^2\over 2M_G} 
+|\bm{\omega}^r|p_2 +\sum_{l=1}^3 \sum_{k+ l> 2}^3{p_kp_l\over 2M_{kl}}\nonumber\\
&+& \epsilon [V_G+V_1(\bm{p})+V_2(\bm{p})+\ldots+V_{r_{opt}}(\bm{p})]\cos\psi_1
+ \epsilon \sum_{q=2}^{q_{opt}}U_q(\bm{p})]\cos(q\psi_1) \\
&+& \epsilon \sum_{\bm{m}} [V_{0,\bm{m}}+V_{1,\bm{m}}(\bm{p})+\ldots
+ V_{20,\bm{m}}(\bm{p})\cos(\bm{m}\cdot\bm{\theta}(\bm{\psi}))~~.\nonumber
\end{eqnarray}
In the above expression:

i) We have already computed $V_G$, while 
\begin{equation}
{1\over M_{kl}}=\sum_{i=1}^3 \sum_{j=1}^3\Upsilon_{ki}{\partial\omega^r_i
\over\partial I_j}\Upsilon_{lj},\qquad
{1\over M_G}\equiv{1\over M_{11}}= \sum_{i=1}^3 
\sum_{j=1}^3 m_{g_i}{\partial\omega^r_i
\over\partial I_j}m_{g_j}~~.
\label{eq25}
\end{equation}
We find the following values:
\begin{eqnarray*}
M_G&=&4.68789151245171606\times 10^{-2},\\
M_{12}^{-1}=M_{21}^{-1}&=&-1.3250433064004110\\
M_{13}^{-1}=M_{31}^{-1}&=&0,\\
M_{22}^{-1}&=&0.95805552130252458,\\
M_{23}^{-1}=M_{32}^{-1}&=&0.19026669807696181\\
M_{33}^{-1}&=&0.81946117880019043.
\end{eqnarray*}

ii) The functions $V_s$ and $V_{s,\bm{m}}$ are homogeneous polynomial of 
degree $s$ in the variables $p_i$, $i=1,2,3$. 

iii) The functions $U_q(\bm{p})$ are polynomial in the variables $p_i$. 

iv) The coefficients $V_{s,\bm{m}}$ are much smaller in size than the 
coefficients $V_s$ or $U_q$, since the former belong to the remainder, 
while the latter belong to the normal form. 

The Hamiltonian (\ref{hamppsi}) does not have precisely the form required 
for the implementation of Chirikov's formulae. We thus proceed in obtaining 
an approximate form of the Hamiltonian, by implementing a number of 
simplifications as follows:

i) By construction, the initial conditions of all the orbits are taken to lie 
nearly exactly on the so called `plane of fast drift' (see section~\ref{sec5}), 
which corresponds to setting initially $p_2\simeq p_3\simeq 0$. Furthermore, 
if we neglect the effect of the remainder, $p_2$ and $p_3$ are preserved 
quantities under the normal form dynamics. We thus set $p_2=p_3=0$ in all 
estimates for the coefficients $V_s$, $U_q$, or $V_{s,\bm{m}}$. 

ii) In contrast, $p_1$ is subject to oscillations in time, since, according 
to Eq.(\ref{hamppsi}), it is a pendulum action variable. For the separatrix 
half-width we find the estimate 
\begin{equation}\label{sephw}
\Delta p_1=2\sqrt{\epsilon |M_G V_G|}\simeq 0.0314 \epsilon^{1/2}~~.
\end{equation}
Since the orbits actually evolve in a thin separatrix-like layer, instead 
of an exact separatrix solution, we may assume that the time evolution 
of $p_1(t)\approx \Delta p_1\cos(\psi_1/2)$ 
does not differ much from the evolution of the momentum along 
an oscillation solution with period $2\pi/\Omega_G$ and amplitude 
$\Delta p_1$. Then, performing the average over one period of motion,
for all odd powers we have $<p_1(t)^n>\simeq 0$, 
while for the even orders we use the approximation:
\begin{equation}\label{p1even}
<p_1(t)^n>\simeq{1\over 2^n}{n!\over (n/2)!(n/2)!}\Delta p_1^n~~.
\end{equation}
Then, we estimate the numerical values of all coefficients 
$V_{s,\bm{m}}(p_1;p_2=p_3=0)$ in Eq.(\ref{hamppsi}), setting 
$V_{s,\bm{m}}(p_1;p_2=p_3=0)=0$ if $s$ is odd, and using the 
expression (\ref{p1even}), with $\Delta p_1$ given by Eq.(\ref{sephw}), 
if $s$ is even. In practice, the remainder contains hundreds of 
thousands of terms, most of which, after the above substitutions, 
are found to be of negligible size. We thus impose a size cut-off limit 
and in subsequent calculations keep only the remainder terms of 
size larger than $10^{-10}$. 

iii) Finally, we ignore all the normal form terms $V_s(\bm{p})$, as 
well as $U_q(\bm{p})$. In fact, one can check that these terms 
introduce corrections of the order $10^{-4}$ of the leading normal 
form term, i.e. $V_G\cos(\psi_1)$. 

With the above simplifications, the Hamiltonian resumes finally 
an approximate form suitable for the implementation of Chirikov's 
formulae, namely
\begin{eqnarray}\label{hamppsi2}
H(\bm{p},\bm{\psi})&=&{p_1^2\over 2M_G} 
+|\bm{\omega}^r|p_2 +\sum_{l=1}^3 \sum_{k+l> 2}^3 {p_kp_l\over 2M_{kl}}
+ \epsilon V_G\cos\psi_1
+ \epsilon \sum_{\bm{m}} \tilde{V}_{\bm{m}}
\cos(\bm{m}\cdot\bm{\theta}(\bm{\psi}))~~
\end{eqnarray}
where the coefficients $\tilde{V}_{\bm{m}}$ have now constant values. 

Under the form (\ref{hamppsi2}), the Hamiltonian lends now itself to 
the computation of Chirikov's diffusion coefficient as follows:

For all amplitudes of the remainder, $\tilde{V}_{\bm{m}}$, we compute the 
coefficients (see appendix~\ref{ap1}) 
\begin{equation}\label{wm}
W_{\bm{m}}=
{4\pi\nu_1(\bm{m})\nu_2(\bm{m})
\tilde V_{\bm{m}}(2|\bm{m\cdot\omega^r}|/\Omega_G)^{2|\xi_{\bm{m}}|-1}
\over
\xi_{\bm{m}}V_G\Gamma(2|\xi_{\bm{m}}|)} 
\exp\left({-\pi |\bm{m\cdot\omega^r}|\over 2\Omega_G}\right)
\end{equation}
where
$$
\nu_k(\bm{m})=\sum_{i=1}^3 m_i \Upsilon_{ik},~~~~k=1,2,3,~~~
\xi_{\bm{m}}=\sum_{k=1}^3 {\nu_k(\bm{m})\over M_{k1}}~~. 
$$
The coefficients $W_{\bm{m}}$ are used in the computation of the per-period 
variation of the pendulum energy integral (see appendix~\ref{ap1}). According 
to Chirikov, we first isolate the perturbing term $\tilde{V}_{\bm{m}}$ 
yielding the largest value of $W_{\bm{m}}$, and we call this term `layer' 
resonance, i.e. the resonance mainly responsible for the formation of 
the thin separatrix chaotic layer at the border of the guiding resonance. 
In our data, we found that the layer resonance is in every case associated 
to the harmonic $\bm{m}_l=(6,-7,-1)$. This amplitude will be hereafter 
denoted by $W_{\bm{l}}$. 

All other terms  $\tilde{V}_{\bm{m}}$ except for the layer one are called 
`driving' resonances. These are the harmonics whose average time variation 
causes a variation of the values of all the normal form integrals. For the 
ten values of $\epsilon$ considered, we computed the ratio of the layer 
resonance amplitude and that of the largest term of the driving resonances. 
These ratios are shown in Table~\ref{tabla1}.

\begin{table}[h!]
\centering
\begin{tabular}{cc}
\hline\hline
\vspace*{-2ex} \\ %
$\epsilon$\hspace{4mm} & $v_{\bm{m}}=W_{\bm{l}}/ W_{\bm{m}}$ \\ %
\hline %
 $0.020$\hspace{4mm} & $7.6$\\%
\hline  %
 $0.018$\hspace{4mm} & $8.1$\\%
\hline  %
 $0.015$\hspace{4mm} & $9.0$\\%
\hline  %
 $0.013$\hspace{4mm} & $9.6$\\%
\hline  %
 $0.012$\hspace{4mm} & $9.9$\\%
\hline  %
 $0.010$\hspace{4mm} & $10.3$\\%
\hline  %
 $0.008$\hspace{4mm} & $10.4$\\%
\hline  %
 $0.007$\hspace{4mm} & $10.2$\\%
\hline  %
 $0.005$\hspace{4mm} & $8.6$\\%
\hline  %
 $0.003$\hspace{4mm} & $2.0$\\%
\hline\hline  \vspace*{-4ex} %
\end{tabular}
\caption{Amplitude of the layer resonance with respect to the
largest driving resonance.}
\label{tabla1}
\end{table}

For all values of $\epsilon$, the largest driving resonance corresponds 
to the harmonic $(6,-4,-3)$,
except for $\epsilon=0.003$, for which the leading driving resonance is $(8,0,-7)$. 
Table~\ref{tabla2} shows the most significant driving resonances for each value of 
$\epsilon$, ordered by the size of the corresponding amplitude $W_{\bm{m}}$.

\begin{table}[h!]
\centering
\begin{tabular}{cc}
\hline\hline
\vspace*{-2ex} \\ %
$\epsilon$\hspace{5mm} & Leading Driving Resonances\\ %
\hline %
 $0.020$\hspace{5mm} & $(6,-4,-3);\, (6,-2,-4);\, (6,-6,-2);\, (4,-1,-3);\, (-2,-2,3);\, (4,-3,-2)$\\%
\hline  %
 $0.018$\hspace{5mm} & $(6,-4,-3);\, (6,-2,-4);\, (4,-1,-3);\, (6,-6,-2);\, (-2,-2,3);\, (4,-3,-2)$\\%
\hline  %
 $0.015$\hspace{5mm} & $(6,-4,-3);\, (6,-2,-4);\, (4,-1,-3);\, (6,-6,-2);\, (-2,-2,3);\, (8,0,-7) $\\%
\hline  %
 $0.013$\hspace{5mm} & $(6,-4,-3);\, (6,-2,-4);\, (4,-1,-3);\, (6,-6,-2);\, (-2,-2,3);\, (8,0,-7)$\\%
\hline  %
 $0.012$\hspace{5mm} & $(6,-4,-3);\, (6,-2,-4);\, (4,-1,-3);\, (6,-6,-2);\, (-2,-2,3);\, (8,0,-7)$\\%
\hline  %
 $0.010$\hspace{5mm} & $(6,-4,-3);\, (6,-2,-4);\, (4,-1,-3);\, (8,0,-7);\, (6,-6,-2);\, (-2,-8,3)$\\%
\hline  %
 $0.008$\hspace{5mm} & $(6,-4,-3);\, (6,-2,-4);\, (4,-1,-3);\, (8,0,-7);\, (-2,-8,3);\, (8,-7,-3)$\\%
\hline  %
 $0.007$\hspace{5mm} & $(6,-4,-3);\, (6,-2,-4);\, (8,0,-7);\, (4,-1,-3);\, (-2,-8,3);\, (8,-7,-3)$\\%
\hline  %
 $0.005$\hspace{5mm} & $(6,-4,-3);\, (6,-2,-4);\, (8,0,-7);\, (-2,-8,3);\, (6,-3,7);\, (4,-1,-3)$\\%
\hline  %
 $0.003$\hspace{5mm} & $(8,0,-7);\, (6,-4,-3);\, (6,-2,-4);\, (10,-3,-7);\, (4,-1,-3);\, (-2,-2,3)$\\%
\hline\hline  \vspace*{-4ex} %
\end{tabular}
\caption{Principal driving resonances ordered by their amplitude $W_{\bm{m}}$
for each $\epsilon$ value.}
\label{tabla2}
\end{table}
The results shown in Table~\ref{tabla1} are in agreement with the assumption
that $W_{\bm{l}}\gg W_{\bm{m}}$ for almost all ${\bm{m}}\ne\bm{l}$. For $\epsilon=0.003$,
we obtain that $W_{\bm{l}}\sim W_{\bm{m}}$, and although Chirikov ~\cite{Ch79} pointed 
out that the approximation $v_{\bm{m}}= W_{\bm{m}}/W_{\bm{l}}\sim 1$ should be 
sufficient to justify all estimates, this seems not to be true. For $\epsilon=0.003$
the amplitudes $W_m$ of three leading driving resonances are very similar to the one
of the layer resonance, so
it seems difficult to distinguish between layer and driving resonances. 
It is interesting to see from Table~\ref{tabla2} 
how certain harmonics increase their importance in the perturbation as $\epsilon$ 
changes from larger to lower values. Notice that the driving resonances $(6,-4,-3); 
(6,-2,-4)$ and $(4,-1,3)$ are always present as leading terms for all values of 
the perturbation parameter. Anyway, the amplitudes of the smaller driving resonances 
shown in Table~\ref{tabla2} are of the order of $\sim 10^{-1}$ or 
$\sim 10^{-2} W_{\bm{m}}$ of the leading one. This difference is even larger for 
$\epsilon=0.003$ where $W_{\bm{m}}$ corresponding to the harmonic $(-2,-2,3)$ is 
$\sim 3\times 10^{-3}$ times the $W_{\bm{m}}$ of $(8,0,-7)$.

Excluding the layer resonance, the scalar diffusion coefficient along the 
direction of the vector ${\bm{\mu}_3}$ can be computed by Chirikov's formula 
\begin{eqnarray}\label{difchifull}
D(\bm{I^r};{\bm{\mu}_3})\approx\frac{2\pi^2F\epsilon^2}
{T_a|\bm{\omega}^r|^2}\sum_{\bm{m}\neq\bm{l}}
{\nu_{3}({\bm{m}})^2\over\nu_{2}({\bm{m}})^2}
\left\vert{2\omega_{\bm{m}}\over\Omega_G}\right\vert^{4|\xi_{\bm{m}}|}
{V^2_{\bm{m}}\over\Gamma^2(2|\xi_{\bm{m}}|)}e^{-{\pi|\omega_{\bm{m}}|
\over\Omega_G}}~~,
\end{eqnarray}
where
$$
T_a(w_s)\approx\frac{1}{\Omega_G}\ln\left(\frac{32e}{w_s}\right),\qquad
T_a\sim \left(\frac{1+\ln{\epsilon}}{\sqrt{\epsilon}}\right)+
\frac{1}{\epsilon}\qquad
\Omega_G(\epsilon)=\sqrt{{\epsilon |V_G|}/{|M_G|}},\qquad
\omega_{\bm{m}}\!=\!\bm{m}\!\cdot\!\bm{\omega}^r\!
=\!\nu_{2}({\bm{m}})|\bm{\omega}^r|,
$$
$\Gamma(x)$ denotes the Gamma function and $w_s$ is the width of the
guiding resonance chaotic layer (see appendix~\ref{ap1}).
In Eq.(\ref{difchifull}), $F$ is a positive constant called the `reduction 
factor', which accounts for correlations between the phases representing 
the initial conditions of the various iterates of the orbits in the 
chaotic layer, as described by the separatrix mapping. The true value of $F$ 
is unknown, but for simplicity we set $F=1$ (see Section~\ref{sec5}). 
{Although Chirikov~\cite{Ch79} suggested, by a crude theoretical estimate, the
value $F\sim 1/3$, we find the introduction of such a ''reduction factor"
rather unnecessary in the regime examined in the present paper, in
which theoretical estimates can be compared to numerical ones only
up to an order of magnitude agreement.

Implementing Eq.(\ref{difchifull}) in our data we obtained the following 
values for $D_C$ in terms of $\epsilon$:
\begin{table}[h!]
\centering
\begin{tabular}{cc}
\hline\hline
\vspace*{-2ex} \\ %
$\epsilon$\hspace{4mm} & $D_C$ \\ %
\hline %
 $0.020$\hspace{4mm} & $8.4\times 10^{-11}$\\%
\hline  %
 $0.018$\hspace{4mm} & $3.0\times10^{-11}$\\%
\hline  %
 $0.015$\hspace{4mm} & $5.0\times10^{-12}$\\%
\hline  %
 $0.013$\hspace{4mm} & $1.3\times10^{-12}$\\%
\hline  %
 $0.012$\hspace{4mm} & $5.9\times10^{-13}$\\%
\hline  %
 $0.010$\hspace{4mm} & $1.2\times10^{-13}$\\%
\hline  %
 $0.008$\hspace{4mm} & $1.9\times10^{-14}$\\%
\hline  %
 $0.007$\hspace{4mm} & $6.9\times10^{-15}$\\%
\hline  %
 $0.005$\hspace{4mm} & $6.7\times10^{-16}$\\%
\hline  %
 $0.003$\hspace{4mm} & $2.4\times10^{-17}$\\%
\hline\hline  \vspace*{-4ex} %
\end{tabular}
\caption{Diffusion coefficient value for $\epsilon\in\mathcal{E}$. 
Note the significant decrease of $D_C$ for $\epsilon=0.003.$ See text.}
\label{table: D_C}
\end{table}

We should emphasize that the obtained values for $D_C$ rest under the assumption 
of normal diffusion, for which the unperturbed (original or transformed) actions 
vary as a linear power of $t$. This is in fact a strong assumption. As discussed 
below, our numerical results indicate that the assumption of normal diffusion 
is an adequate first approximation. However, there are additional features in 
all our obtained diffusion curves, whose quantitative description goes beyond 
the assumptions of Chirikov's theory. At any rate, as it will be shown in the next 
section, the diffusion coefficients computed numerically under the assumption 
of normal diffusion agree very well with the diffusion coefficients reported 
in table \ref{table: D_C}. To this comparison we now turn our attention. 

\section{Numerical results. Comparison of all estimates}\label{sec5}
In the present section, we will present the results of numerical simulations 
of ensembles of orbits in our system, aiming to determine the diffusion 
coefficient $D$ along the simple resonance, as a function of $\epsilon$, 
in a purely numerical way. In what follows, we present computations using 
two different sets of variables, namely the canonical variables in the 
original Hamiltonian, and those arising after an optimal canonical transformation. 
To separate in notation the former from the latter, we use the superscript $(0)$ 
to denote the old canonical variables, i.e., the angles  $\bm{\theta}^{(0)}$, and 
the actions $\bm{ I^{(0)}}$ (or $\bm{ p^{(0)}}$ in Chirikov's basis). We use bar, 
or non-bar, symbols to denote all ensemble quantities defined over the set 
of the old, or the new, action variables respectively.  

\subsection{Statistical quantities}

\label{subsec: statistical_quant}
Using the results from the integration of ensembles of orbits, we measure 
the time variation of the {\it variance} of the old and new action variables, 
$p_i^{(0)}$ and $p_i$ respectively, for $i=1,2,3.$ The statistical quantities 
presented below are {\it ensemble} averages computed numerically over a finite 
number, $N_p$, of test particles. The initial conditions of these test particles 
are chosen in a small domain in the action space, all of them having the same 
total energy. 

Let $p_j(t,i)$ be the value of the $j$--th component of the vector $\bosy{p}$,
at the time $t$, associated to the $i$--th particle.
Thus, the (instantaneous) mean value of this component is given by:
\begin{equation*}
  \mu_j(t)\equiv \langle p_j(t) \rangle \equiv \frac{1}{N_p}\sum_{i=1}^{N_p} p_j(t,i), 
\end{equation*}
and the corresponding variance is: 
\begin{equation*}
  \label{intro_variance_vector}
  \sigma_j^2(t) \equiv \langle ( p_j(t)-\mu_j(t) )^2 \rangle 
\equiv \frac{1}{N_p} \sum_{i=1}^{N_p}(p_j(t,i)-\mu_j(t))^2.
\end{equation*}
Analogously, we can define the mean and the variance of the old actions:
\begin{equation}
  \label{intro_old_variance}
        {\bar{\mu}}_j(t)\equiv \langle p_j^{(0)}(t) \rangle, 
        \qquad
        {{\bar{\sigma}}_j^2}(t) \equiv \langle ( p_j^{(0)}(t)
-{\bar{\mu}}_j(t) )^2 \rangle.
\end{equation}

Diffusion processes are commonly characterized by a power law relationship 
of the form $\sigma^2(t)=c~t^\eta$, with $c>0$. If $\eta=1$ we have normal 
diffusion, while in case of $\eta<1$, the phenomenon is called  subdiffusion, 
or when $\eta>1$ it is called superdiffusion. In the normal diffusion case, 
within some time interval $[t_0,t_f]$, it is possible to define a numerical 
diffusion coefficient, $D$. In this work, the diffusion coefficient is 
associated to $\sigma_3^2$ through a least square fit of the ansatz:
\begin{equation}
  \label{intro_dif_coef_variance}
  \sigma_3^2(t) =  D t + \rho,
\end{equation}
where $D$ and $\rho$ are the fitted parameters.

\subsection{The ensembles}
\label{subsec: ensembles}

For each value of $\epsilon\in\mathcal{E}$, we consider one ensemble of 
$N_p=10^3$ particles. The ICs of the ensembles are 
chosen inside the chaotic layer of the guiding resonance. More specifically, 
they are located in a neighborhood of the unperturbed separatrix of the simple 
pendulum model associated to the resonant action ($\boldsymbol{I}^r$). All the 
ICs belong to the same plane used to compute the SALI maps (as in 
Fig.\ref{guiding}), i.e. they satisfy $\theta_i^{(0)}=\pi/2$ ($i=1,2,3$) 
and $H(\bosy{I}^{(0)},\bosy{\theta}^{(0)})=h$ so that they are solved from:
$H_0(\bosy{I}^{(0)})\equiv H(\bosy{I}^{(0)},\pi/2,\pi/2,\pi/2)=h$.
They are selected at random inside a square of size $4\times 10^{-12}$.

Fig.~\ref{fig: SALI_ICs} display
the SALI map in a neighborhood of  $\boldsymbol{I}^r$ together with the location
of the initial ensemble for $\epsilon$ equal to $0.015$ and $0.003$, respectively.
The left panel of Fig.~\ref{fig: SALI_ICs} shows a magnification in the neighborhood
of the ensembles of ICs (green square) for the smallest perturbation parameter. 
\begin{figure}[ht!]  
\begin{tabular}{rcl}      
\hspace{-5mm}      
\includegraphics[width=0.38\textwidth]{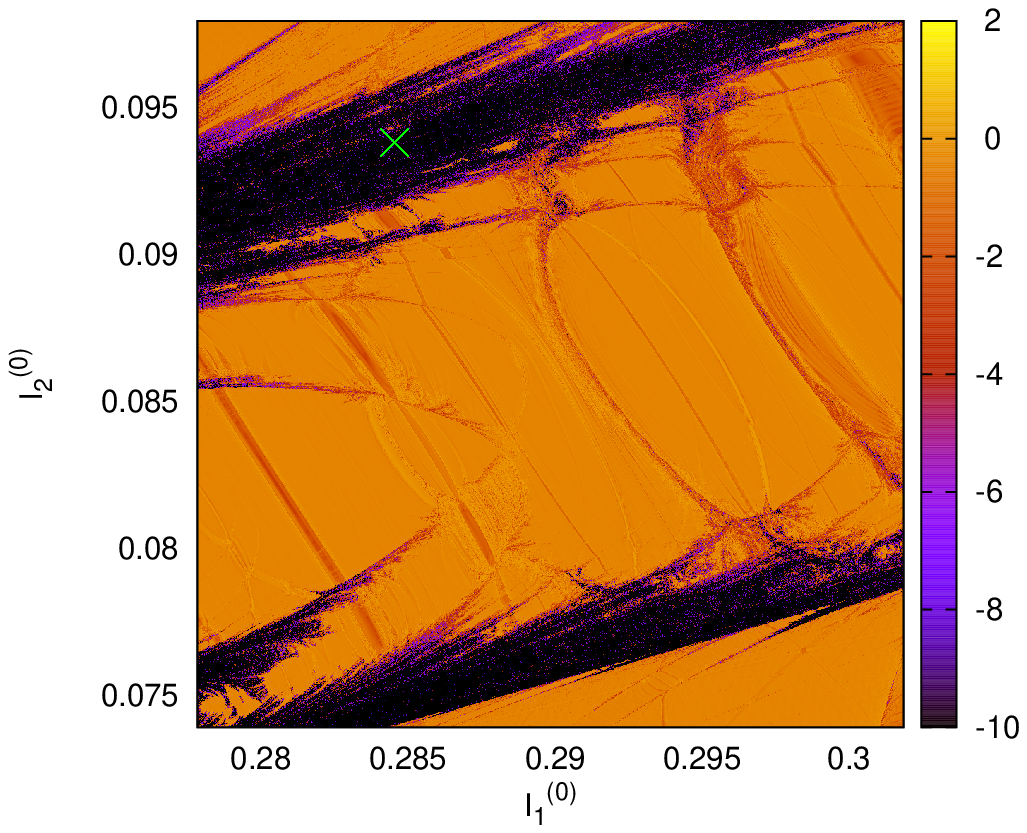}
\hspace{-12.5mm}
      \includegraphics[width=0.38\textwidth]{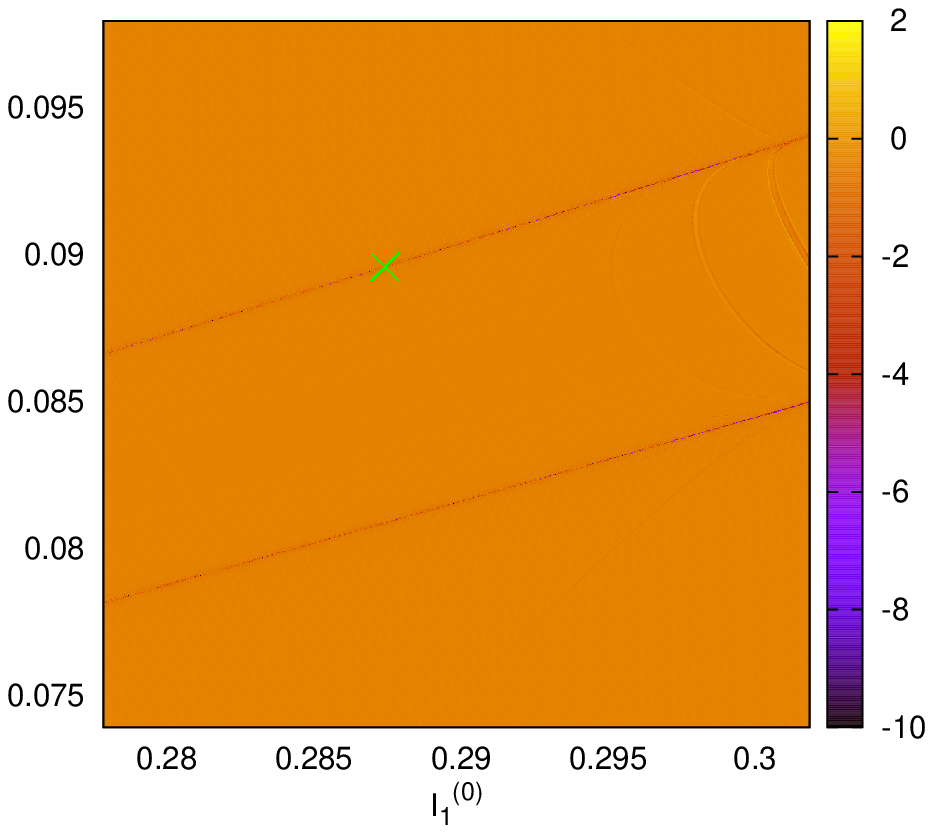}
\hspace{-10.5mm}
      \includegraphics[width=0.38\textwidth]{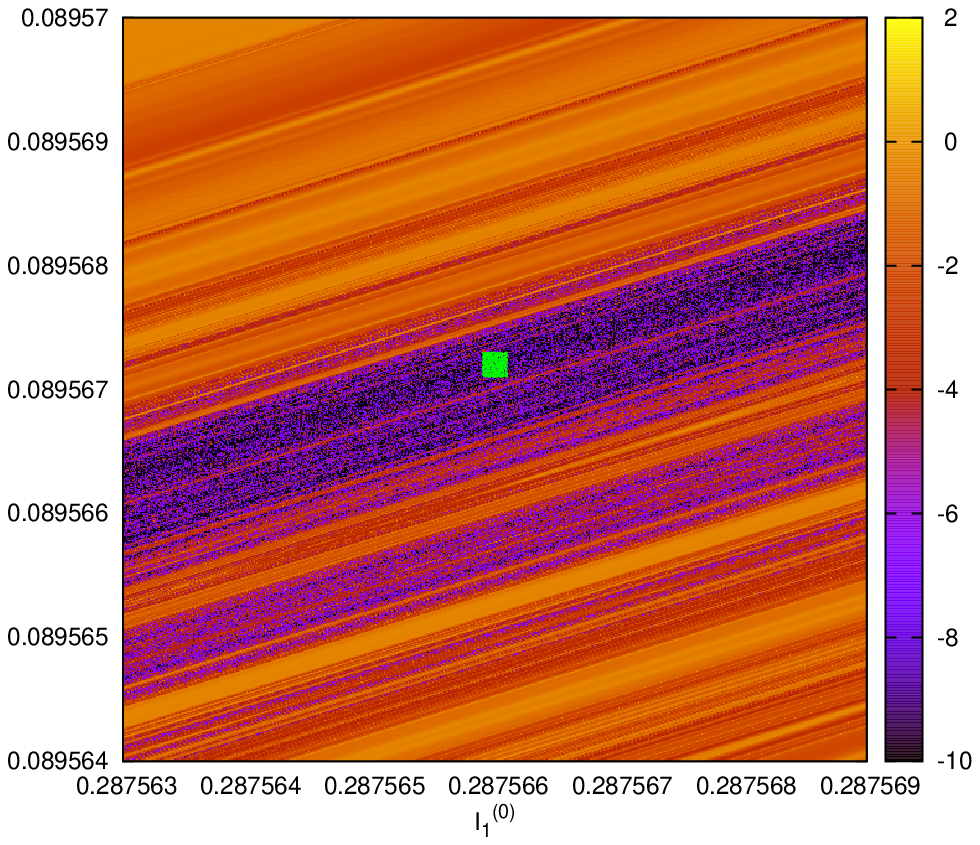}
\end{tabular}
  \caption{Location of the ICs for $\epsilon=0.015$~(left panel)  and 
    $\epsilon=0.003$~(middle and right panels). The green square in the third plot 
    shows the size and shape of the ensemble.}%
  \label{fig: SALI_ICs}
\end{figure}

\subsection{The measurements}
\label{subsec: measurements}
The numerical integrations of the trajectories were performed with a $8^{th}$ 
order symplectic integrator called S8b and elaborated by Teloy, Freiburg, as mentioned 
in ~\cite{SS00}. We use a double precision arithmetic in cartesian variables considering 
a time step $\Delta t_{int}=10^{-2}$. For all $\epsilon\in{\cal E}$, the ensembles were 
numerically evolved up to the time $10^7$. 

Before computing the statistical parameters for all the $\epsilon$ values, we 
qualitatively show the dynamics of these ensembles in the original action space, 
for $\epsilon=0.012$ and $\epsilon=0.015$. We use a double section technique,
applied in \cite{LGF03} among others. While integrating the test particles 
we consider the surface of section $x_1=0$ ($y_1>0$) and collect those points at which 
the intersecting orbit also satisfies the condition $x_2^2+x_3^2\leq \delta^2$ 
($y_2>0$, $y_3>0$), with $\delta=0.002$. In terms of angle variables, the double
section is equivalent to $\theta_1=\pi/2$ with both $\theta_2$ and  $\theta_3$ 
belonging to a certain neighborhood of $\pi/2$, whose size decays to zero with 
$\delta$.

In Fig.~\ref{fig: SOAS_eps0.012} we plot, for the smaller perturbation parameter, 
all the intersections with the double section starting at $t=0$ up to four 
final times: $t=10^5,~5\times10^5,10^6,~5\times10^6$.

\begin{figure}[b!]
  \begin{center}
    \begin{tabular}{rl}
     \fcolorbox{white}{white}{\includegraphics[width=0.35\textwidth]{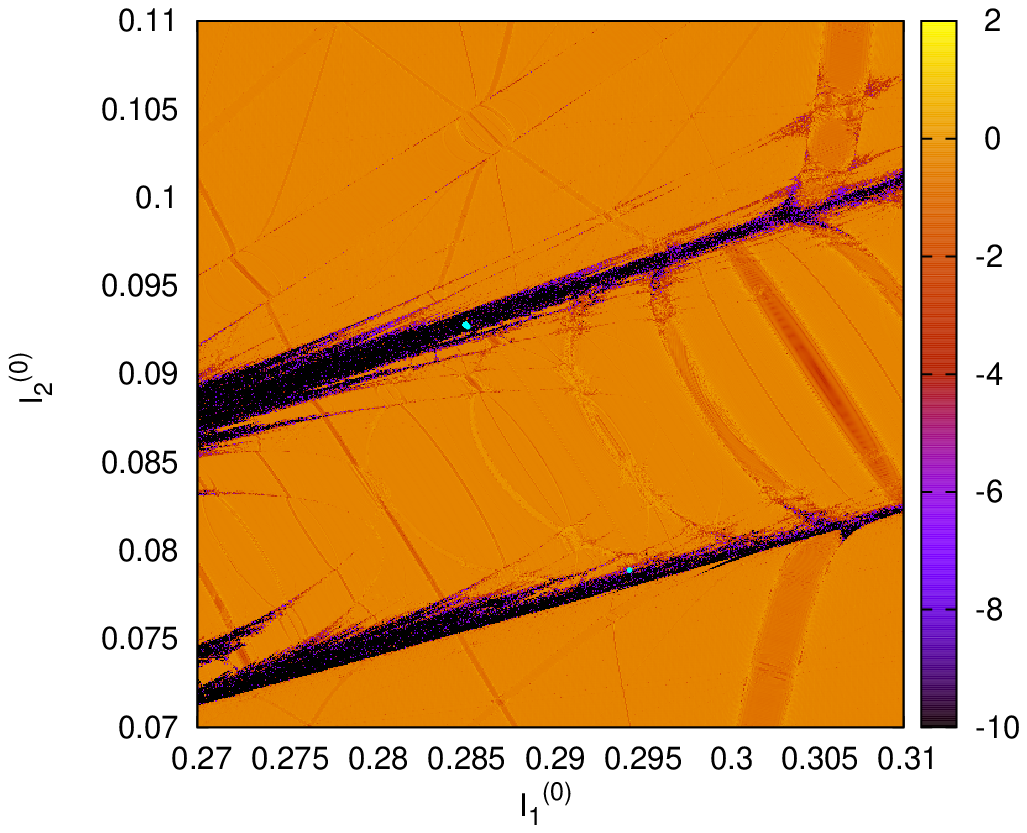}}&
     \hspace{-5mm}\fcolorbox{white}{white}{\includegraphics[width=0.35\textwidth]{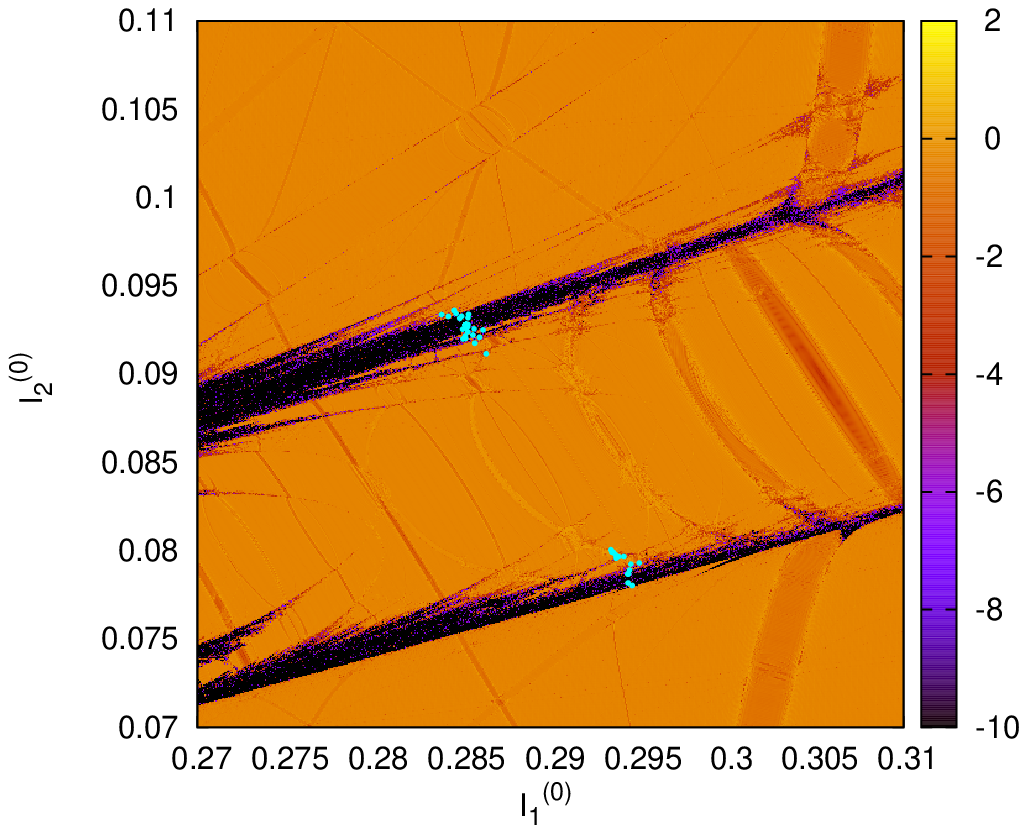}}\\
      \fcolorbox{white}{white}{\includegraphics[width=0.35\textwidth]{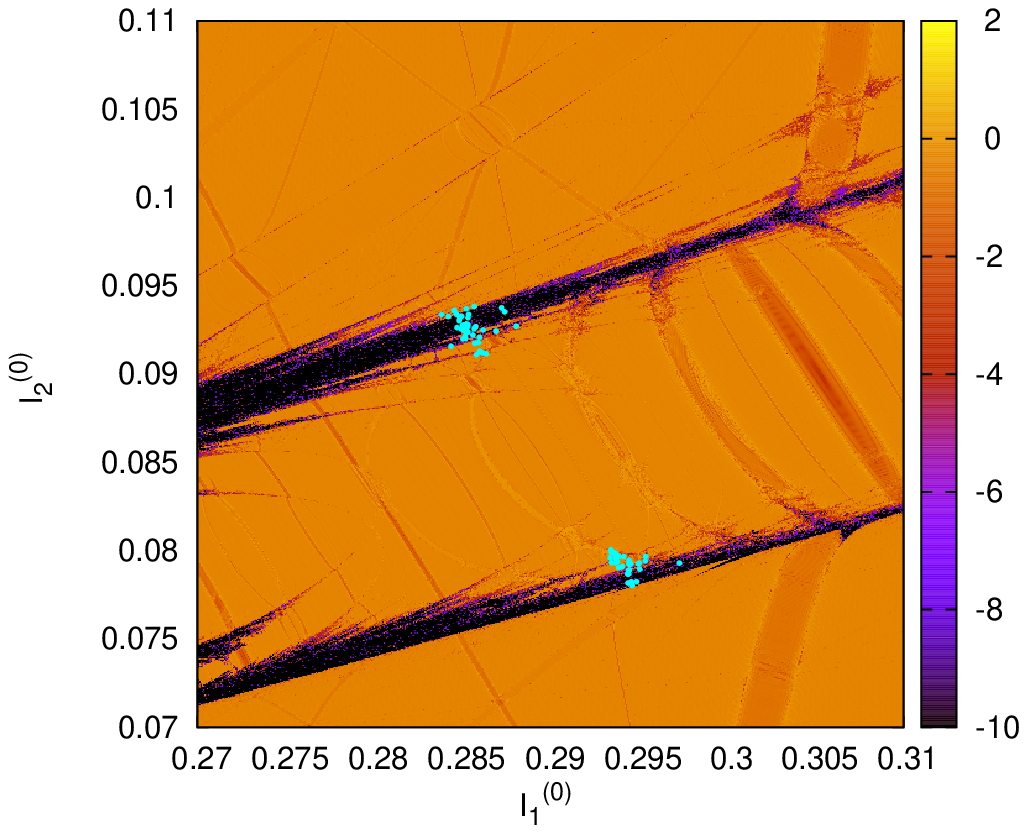}}&
       \hspace{-5mm}\fcolorbox{white}{white}{\includegraphics[width=0.35\textwidth]{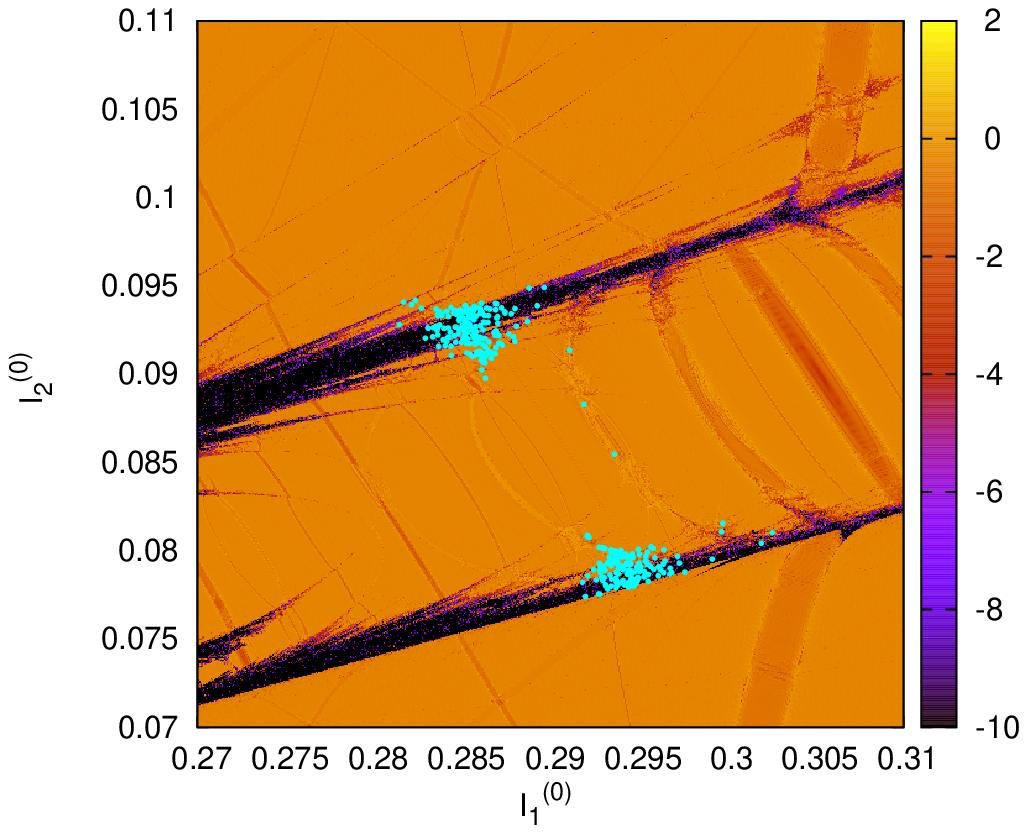}}
   \end{tabular}
    \caption{Intersection of the trajectories with the double
      section defined by $x_1=0$, $(x_2^2+x_3^2)^{1/2}\leq 0.002$ 
      and $y_i>0$ ($i=1,2,3$), 
      projected onto the $[I_1,I_2]$ plane for times:
      $t\leq 10^5$ (top--left), $t\leq 5\times 10^5$ (top--right), 
      $t\leq 10^6$ (bottom--left) and $t\leq 5\times 10^6$ (bottom--right). 
      The data corresponds to $\epsilon=0.012$.}
      \label{fig: SOAS_eps0.012}
  \end{center}
\end{figure}

Similarly, in Fig.~\ref{fig: SOAS_eps0.015} we plot, for a larger perturbation parameter, 
all the intersections with the double section. There we can see how the ensemble expands 
along the  stochastic layer of the guiding resonance. 
\begin{figure}[ht!]
  \begin{center}
    \begin{tabular}{rl}
      \fcolorbox{white}{white}{\includegraphics[width=0.35\textwidth]{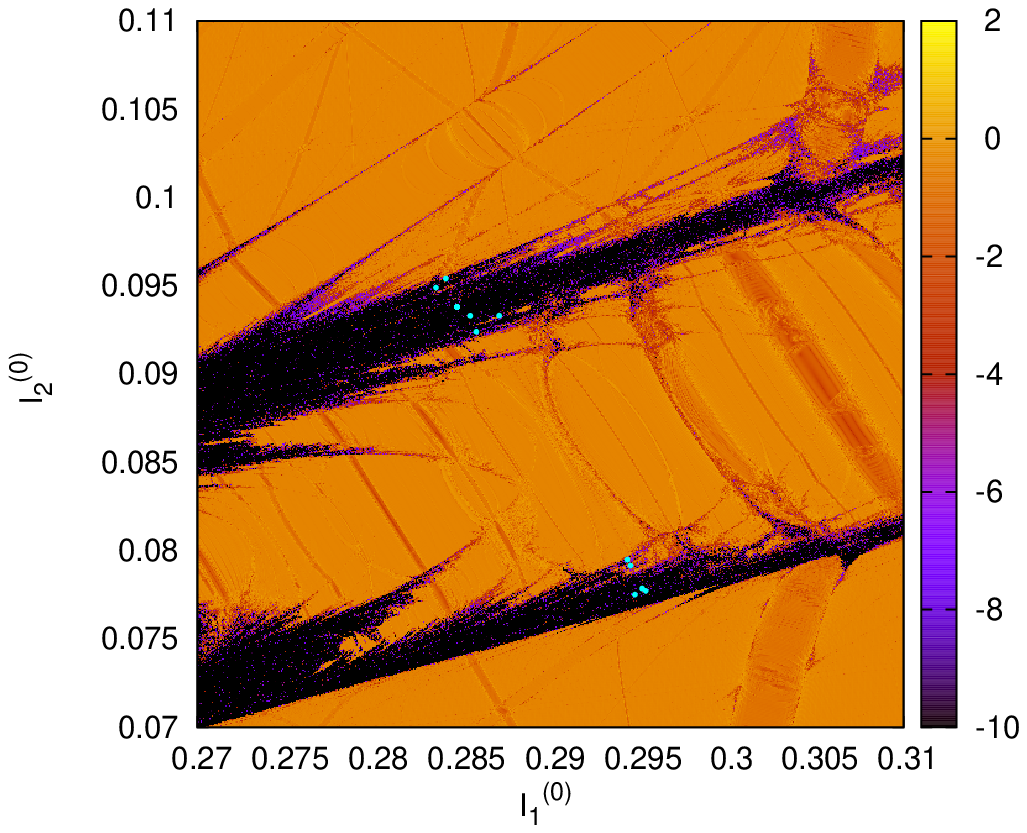}}&
      \hspace{-5mm}\fcolorbox{white}{white}{\includegraphics[width=0.35\textwidth]{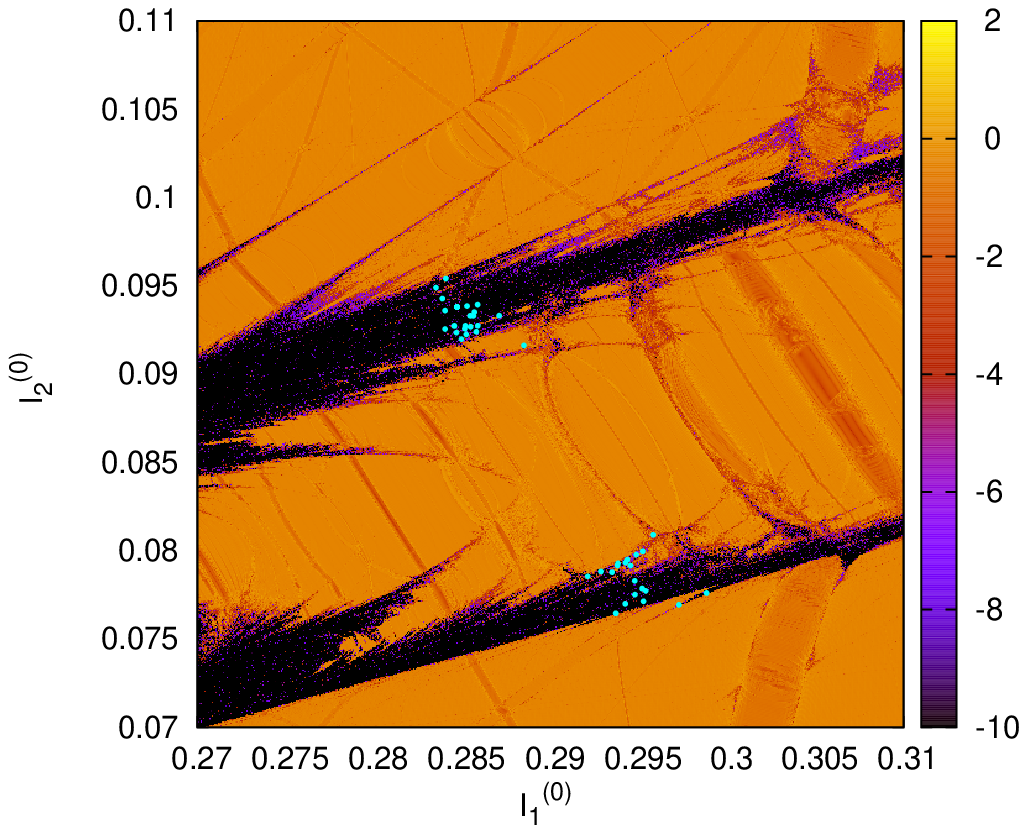}}\\
       \fcolorbox{white}{white}{\includegraphics[width=0.35\textwidth]{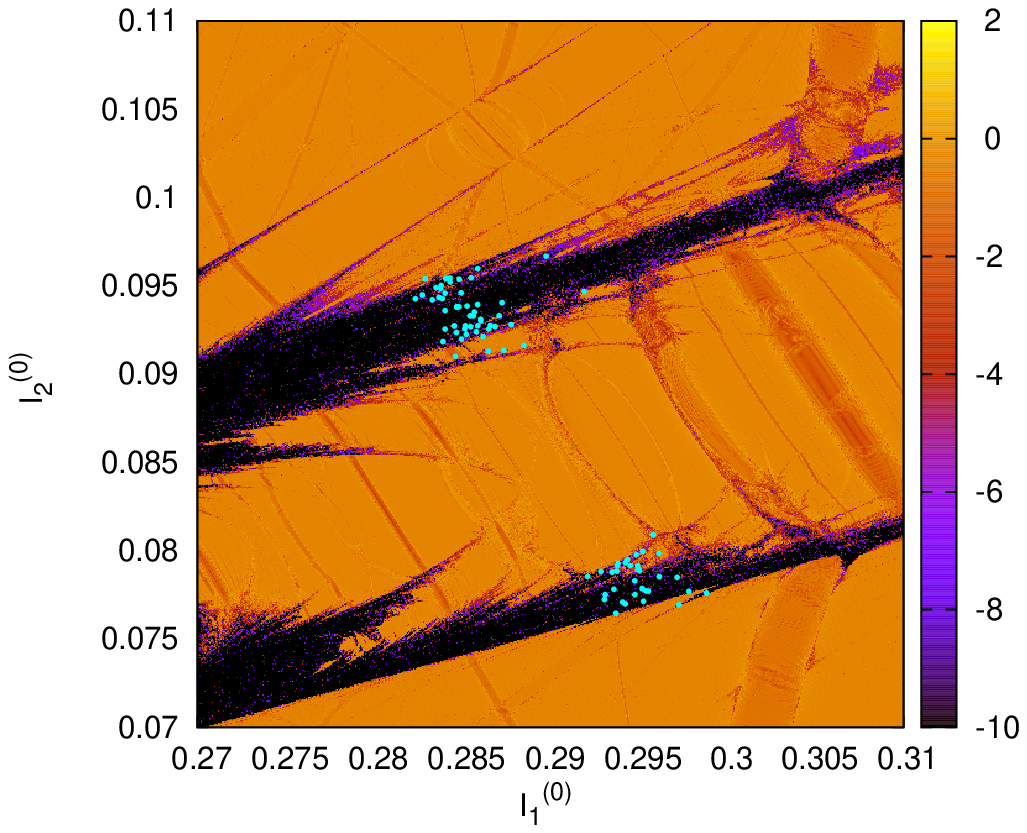}}&
        \hspace{-5mm}\fcolorbox{white}{white}{\includegraphics[width=0.35\textwidth]{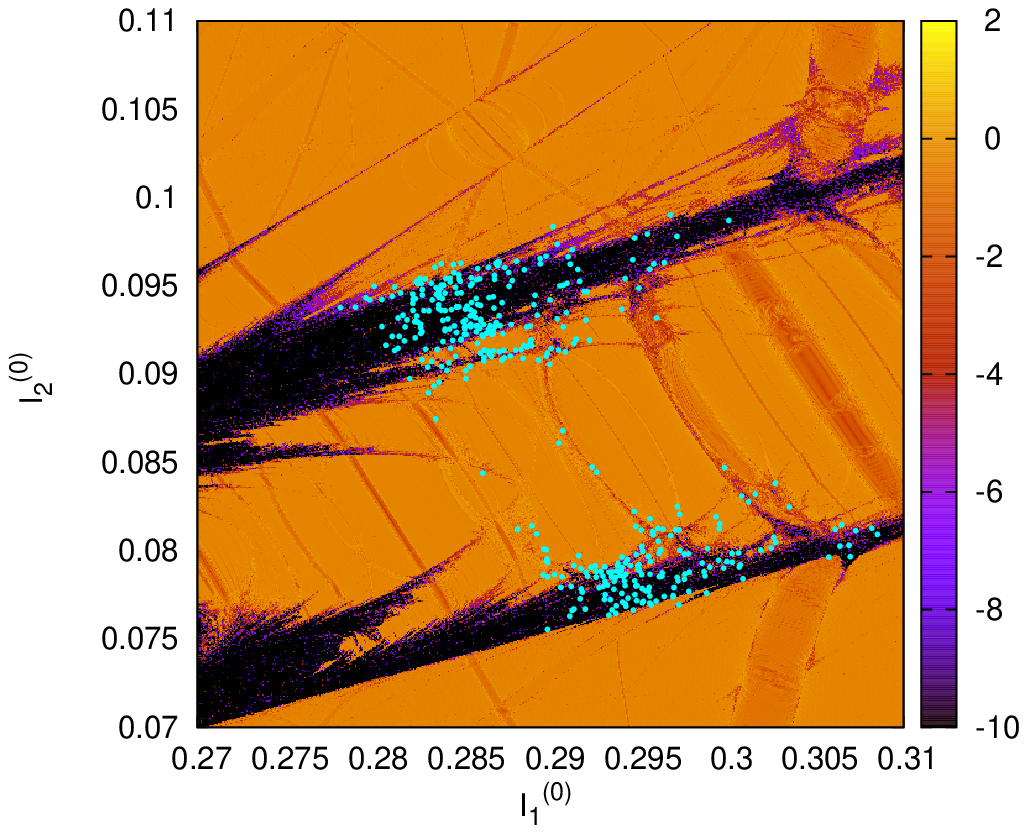}}
    \end{tabular}
    \caption{Intersection of the trajectories with the double
      section defined by $x_1=0$, $(x_2^2+x_3^2)^{1/2}\leq 0.002$ 
      and $y_i>0$ ($i=1,2,3$), 
      projected onto the $[I_1,I_2]$ plane for times:
      $t\leq 10^5$ (top--left), $t\leq 5\times 10^5$ (top--right), 
      $t\leq 10^6$ (bottom--left) and $t\leq 5\times 10^6$ (bottom--right). 
      The data corresponds to $\epsilon=0.015$.}
      \label{fig: SOAS_eps0.015}
  \end{center}
\end{figure}

For the whole set $\mathcal{E}$, we have computed the evolution of the variance of the
three components of $\bosy{p}^{(0)}$, as given in Eq.~(\ref{intro_old_variance}). 
The left panel of Fig.~\ref{fig: old_variance_three_directions_eps0.015-0.005} 
displays ${{\bar{\sigma}}_j^2}(t)$ for $j=1,2,3$, for $\epsilon=0.015$ in colors red, 
green and blue, respectively. We can see that both ${{\bar{\sigma}}_1^2}(t)$  and 
${{\bar{\sigma}}_2^2}(t)$ are bounded quantities within this time interval, 
as expected. We also notice that, on average, ${{\bar{\sigma}}_3^2}(t)$ has 
a secular growth. 
The initial variance is ${{\bar{\sigma}}_3^2}(0)\approx 7.2\times10^{-15}$ 
and has a relatively large jump that starts at $t\approx 900$. 
The behavior for $\epsilon=0.005$ is shown in the right panel of
Fig.~\ref{fig: old_variance_three_directions_eps0.015-0.005}. For the rest of the 
$\epsilon$ values, the observed behaviors are qualitatively similar to the above 
exposed ones.
\begin{figure}[ht!]
  \begin{center}
    \begin{tabular}{rl}     
      \fcolorbox{white}{white}{\includegraphics[width=0.35\textwidth]{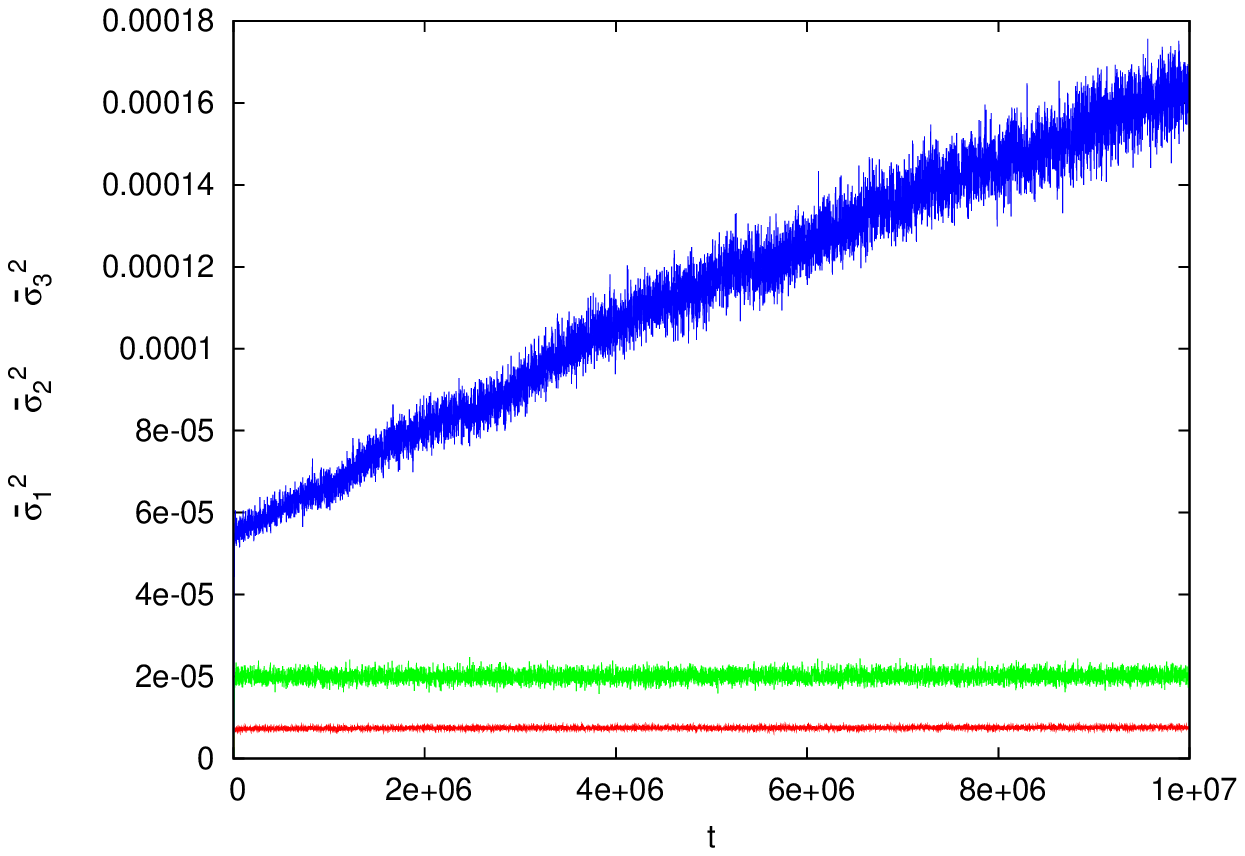}}
     \fcolorbox{white}{white}{\includegraphics[width=0.35\textwidth]{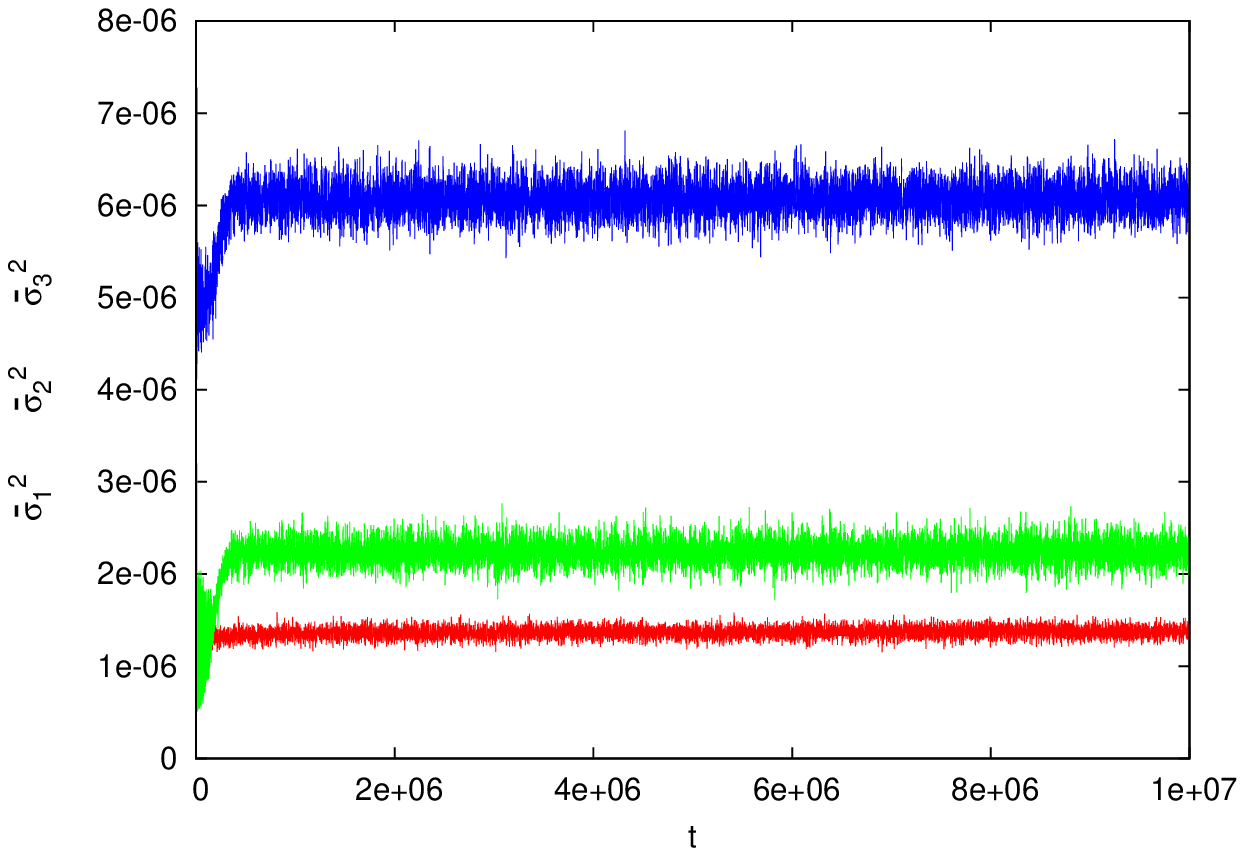}}
    \end{tabular}
    \caption{{The left panel displays the 
      variance evolution in the original actions for $\epsilon=0.015$ along the three directions of
      the new basis: ${{\bar{\sigma}}_j^2}(t)$, for $j=1,2,3$, 
      in colors red, green and blue, respectively.
      ${{\bar{\sigma}}_1^2}(t)$  and ${{\bar{\sigma}}_2^2}(t)$  are bounded 
      quantities within this time interval while ${{\bar{\sigma}}_3^2}(t)$ presents a 
      secular growth.
      The right panel displays, for $\epsilon=0.005$, the values of
      ${{\bar{\sigma}}_j^2}(t)$, for $j=1,2,3$, 
      in colors red, green and blue, respectively. The three quantities appear to be bounded in
      the considered time interval.}
    }
      \label{fig: old_variance_three_directions_eps0.015-0.005}
  \end{center}
\end{figure}

From now on we work only with the variance in the $\bm{\mu}_3$ direction, comparing 
its time evolution as computed using the original action variables or the ones 
corresponding to the optimal canonical transformation. 
Figs.~\ref{fig: old_new_eps0.015},~\ref{fig: old_new_eps0.010} and~\ref{fig: old_new_eps0.005}
show, respectively for $\epsilon=0.015$, $0.010$ and  $0.005$, the values of  
${\bar{\sigma}}_3^2(t)$, in blue, and $\sigma_3^2(t)$, in black. Figs.~\ref{fig: new_eps0.015},
~\ref{fig: new_eps0.010} and~\ref{fig: new_eps0.005} show only the value of  $\sigma_3^2(t)$ 
for the same perturbation parameters.

\begin{figure}[ht!]  
  \begin{center}
    \subfigure[$\epsilon= 0.015$]{%
      \label{fig: old_new_eps0.015}
      \includegraphics[width=0.3\textwidth]{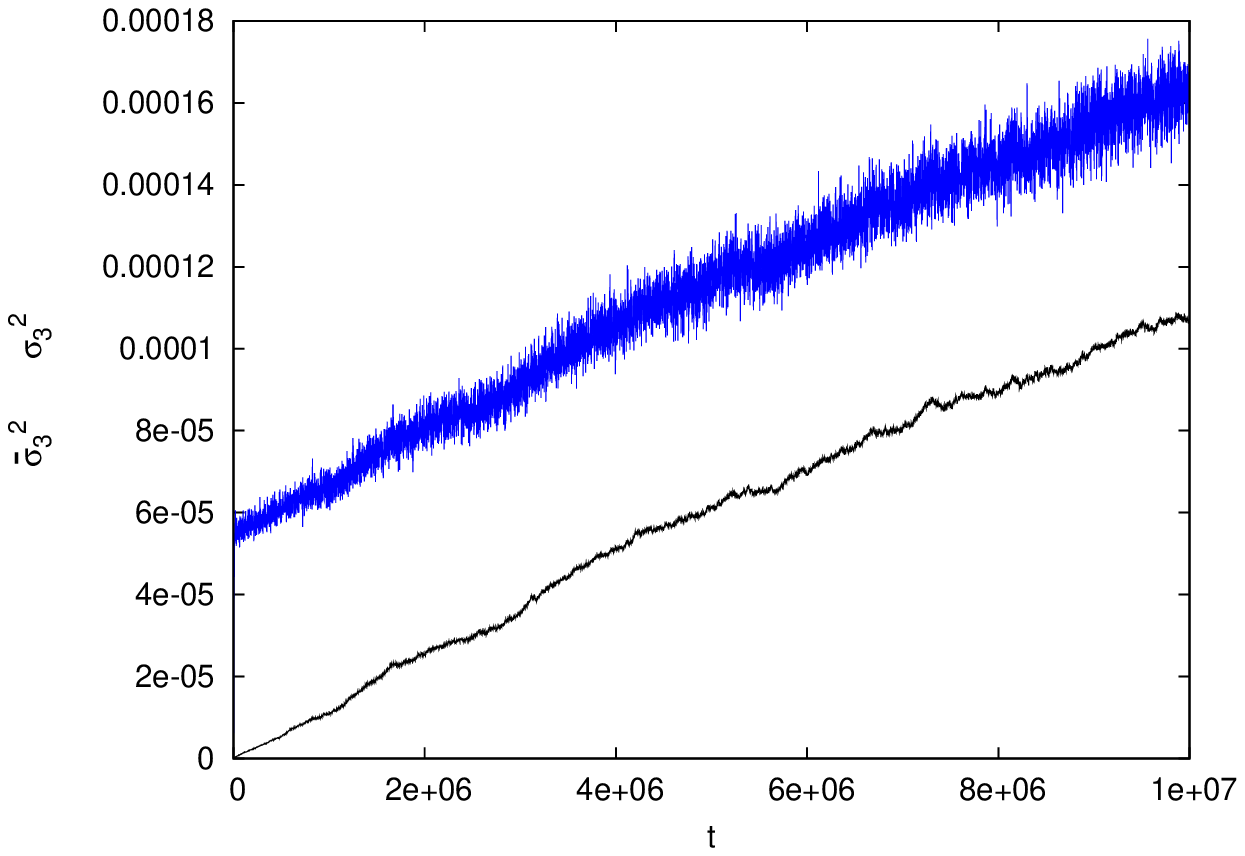}
    }
    \subfigure[$\epsilon= 0.015$]{%
      \label{fig: new_eps0.015}
      \includegraphics[width=0.3\textwidth]{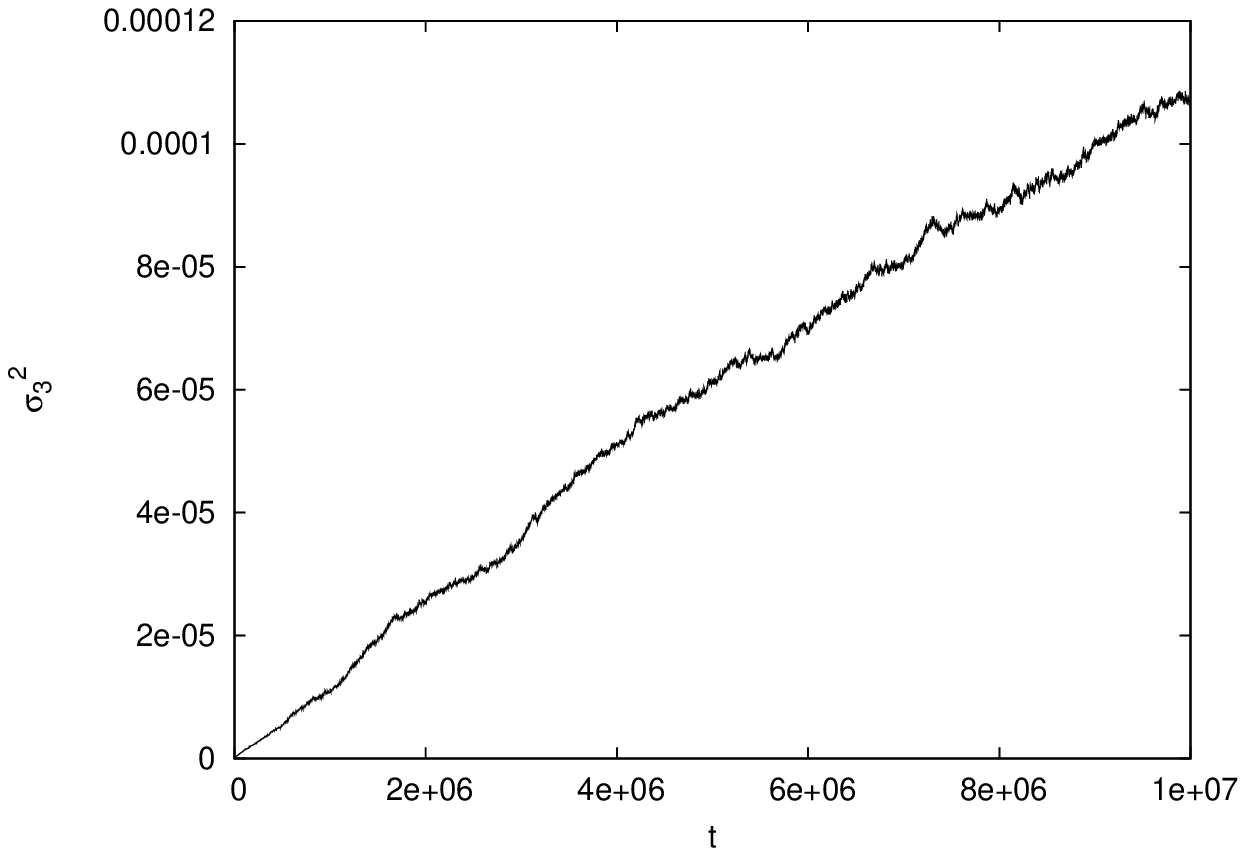}
    }\\
    \subfigure[$\epsilon= 0.010$]{%
      \label{fig: old_new_eps0.010}
      \includegraphics[width=0.3\textwidth]{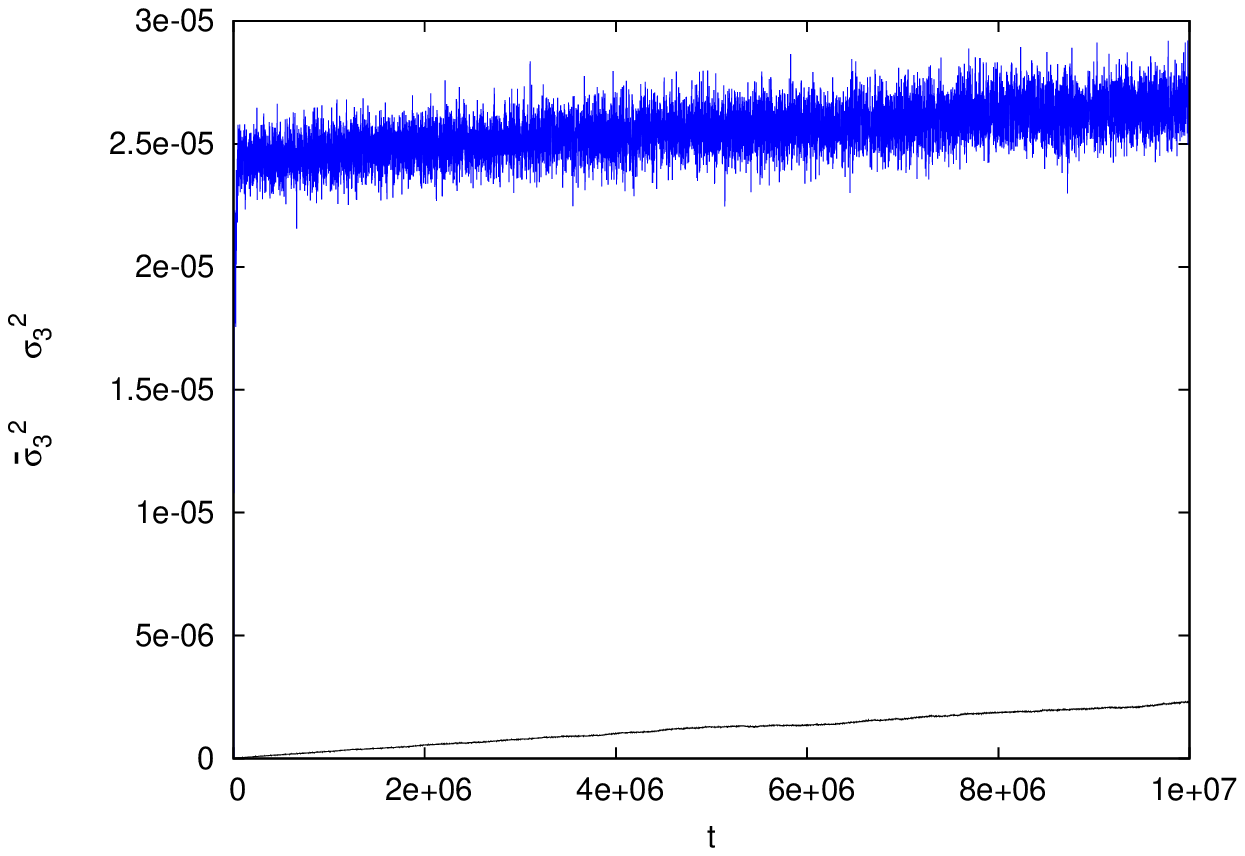}
    }%
    \subfigure[$\epsilon= 0.010$]{%
      \label{fig: new_eps0.010}
      \includegraphics[width=0.3\textwidth]{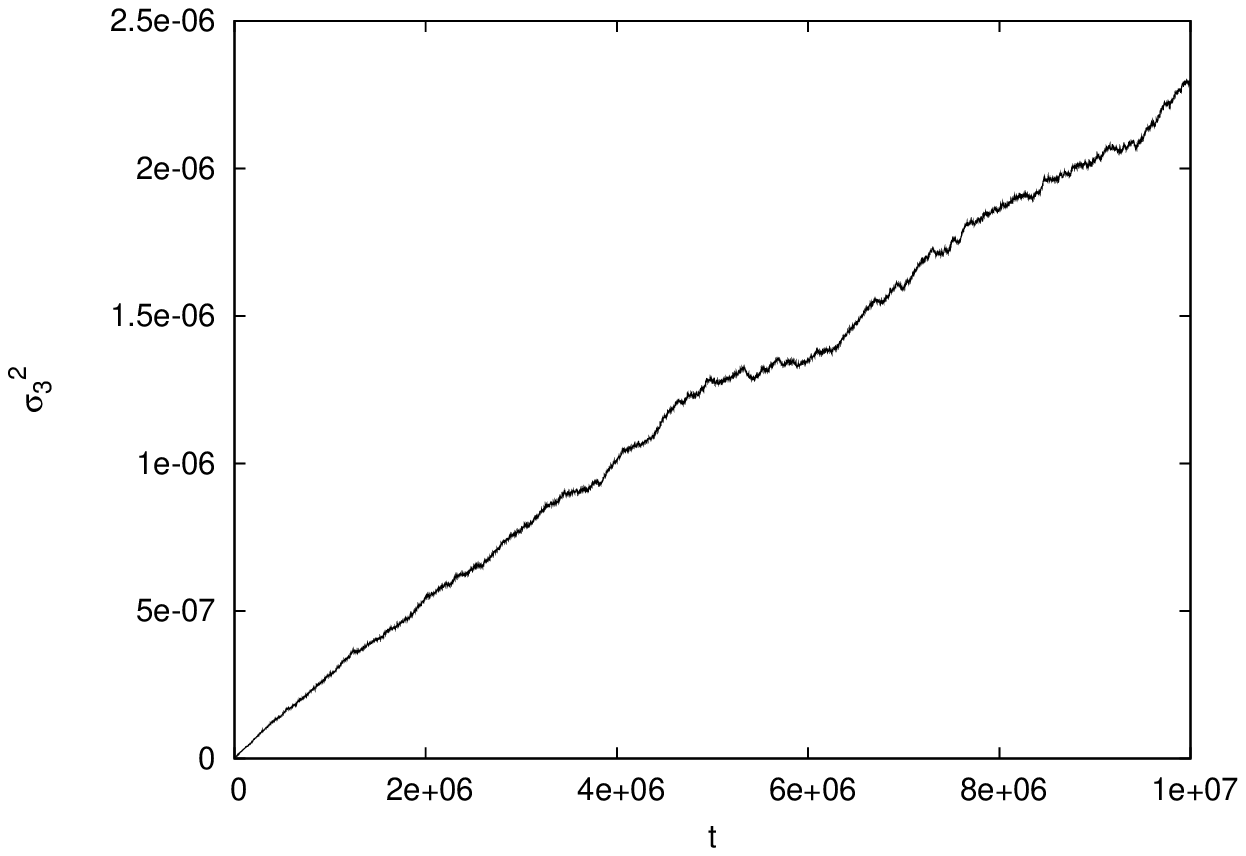}
    }\\
    \subfigure[$\epsilon= 0.005$]{%
      \label{fig: old_new_eps0.005}
      \includegraphics[width=0.3\textwidth]{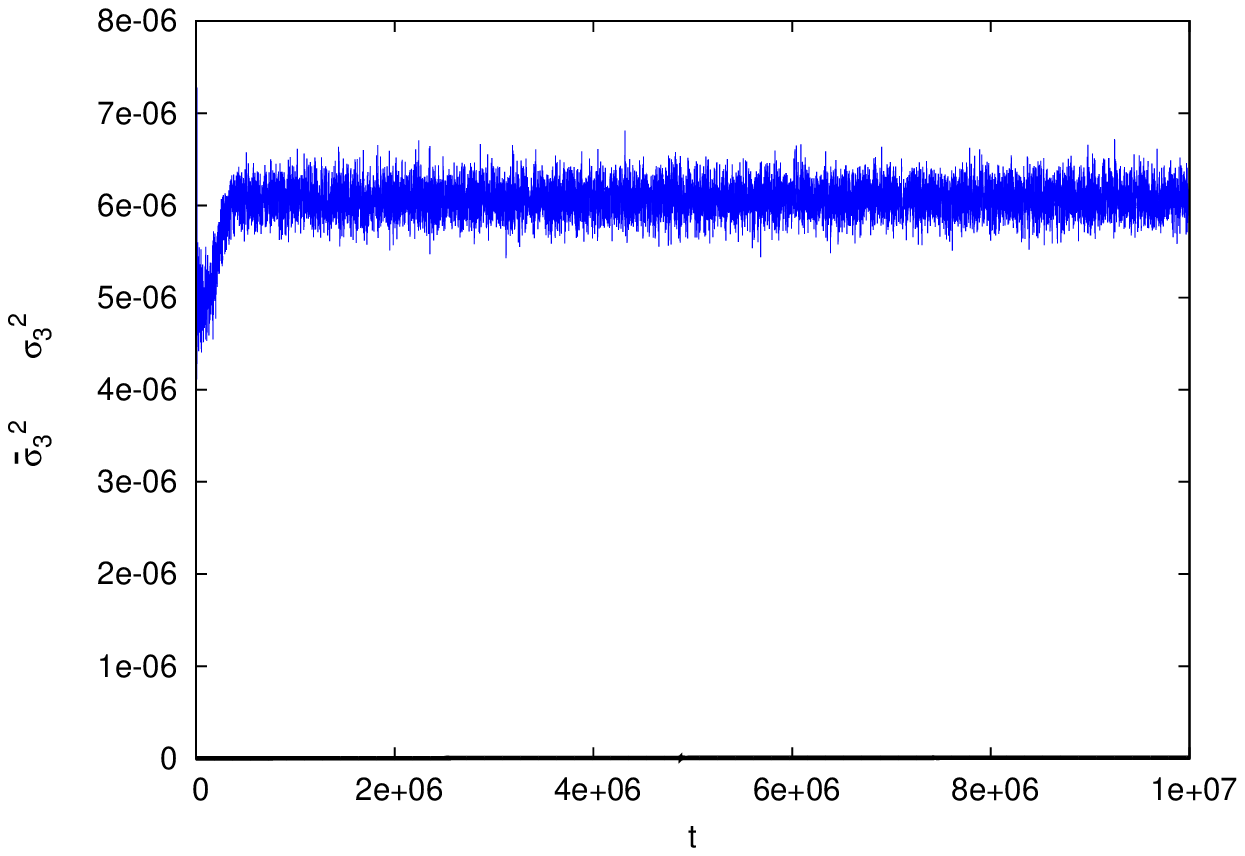}%
    }
    \subfigure[$\epsilon= 0.005$]{%
      \label{fig: new_eps0.005}
      \includegraphics[width=0.3\textwidth]{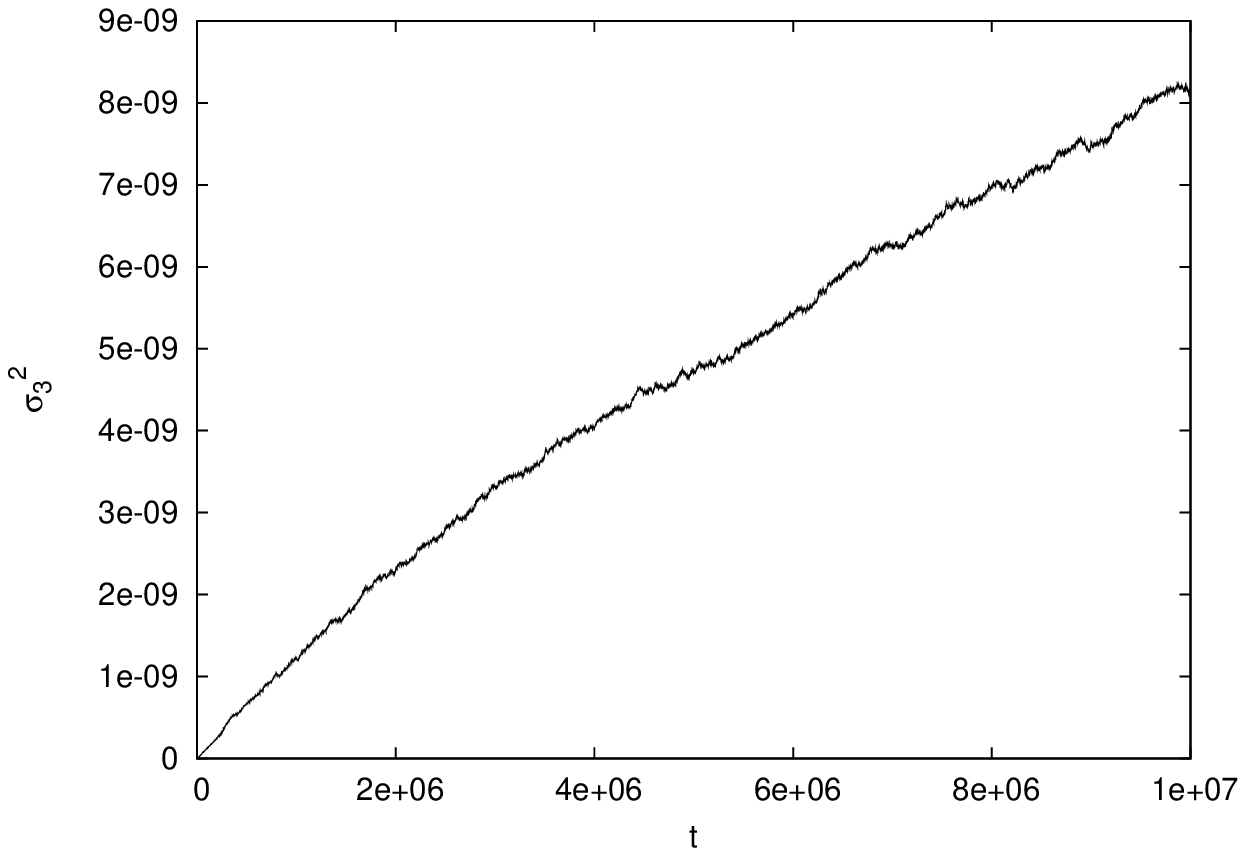}
    }
\end{center}
  \caption{Variance evolution in the original and optimal actions for 
    $\epsilon= 0.015$, $0.010$ and  $0.005$.
    The left hand plots display both ${\bar{\sigma}}_3^2(t)$ and $\sigma_3^2(t)$, 
    in colors blue and black, respectively. 
    The right hand plots display only $\sigma_3^2(t)$.
    We observe a transition in which diffusion can be measured in both families of
    actions obtaining the same average result ($\epsilon=0.015$), 
    towards a state in which diffusion is undetectable in the old actions but precisely
    measurable in the correct ones ($\epsilon=0.005$). The case of $\epsilon=0.010$ is qualitatively
    similar to the one for $\epsilon=0.015$ but is close to the limit of 
undetectability 
    in the old variables.}%
  \label{fig: old_new_eps0.005_0.010_0.015}
\end{figure}

In Fig.~\ref{fig: old_new_eps0.015} the main observation is that the variance 
${\bar{\sigma}}_3^2$ computed in the original canonical variables exhibits 
significant fluctuations, while $\sigma_3^2$ (in the new canonical variables) 
evolves in a smooth linear way without any initial jump. Despite this difference, 
it is remarkable that, for $\epsilon=0.015$, both quantities exhibit a quite similar 
average slope in time.
Now let us focus on $\epsilon=0.010$ (Fig.~\ref{fig: old_new_eps0.010}). 
We see that  ${\bar{\sigma}}_3^2$ still has a measurable average slope and
equal to the one of $\sigma_3^2$. However, the relative difference between 
the two quantities is higher than in the previous case. In fact, the fluctuations 
in ${\bar{\sigma}}_3^2$ are such that the slow systematic time variation is 
just visible in these variables, up to $t=10^7$. 

This picture changes dramatically for smaller values of $\epsilon$, like $0.005$. 
Inspecting the right panel of Fig.~\ref{fig: old_variance_three_directions_eps0.015-0.005},
we cannot distinguish any measurable secular growth in ${\bar{\sigma}}_3^2$. In 
fact, the growth exists, but it is completely ``hidden'' by the large variations 
due to the `deformation' effects. Thus, in practice the diffusion 
rate associated to $p_3^{(0)}$ cannot be experimentally measured. 
Figs.~\ref{fig: old_new_eps0.005} and~\ref{fig: new_eps0.005} show that 
the variance in the new variable $p_3$ is many orders of magnitude smaller 
than ${\bar{\sigma}}_3^2$. Furthermore, all fluctuations due to 
deformation effects are absorbed by the normalizing transformation, 
and do not show up in the time evolution of $\sigma_3^2$. 
This allows to identify and measure the diffusion rate, in this case 
using {\it only} the new canonical variables.

We have not introduced yet any assumption about the dependence of 
$\sigma_3^2(t)$ on time. The closeness to normal diffusion can be graphically 
estimated in Fig.~\ref{fig: Variance_evolution_Log}, displaying the values 
of $\sigma_3^2(t)$ in logarithmic scale $\forall\epsilon\in\mathcal{E}$.
\begin{figure}[ht!]
  \begin{center}
      \fcolorbox{white}{white}{\includegraphics[width=0.48\textwidth]{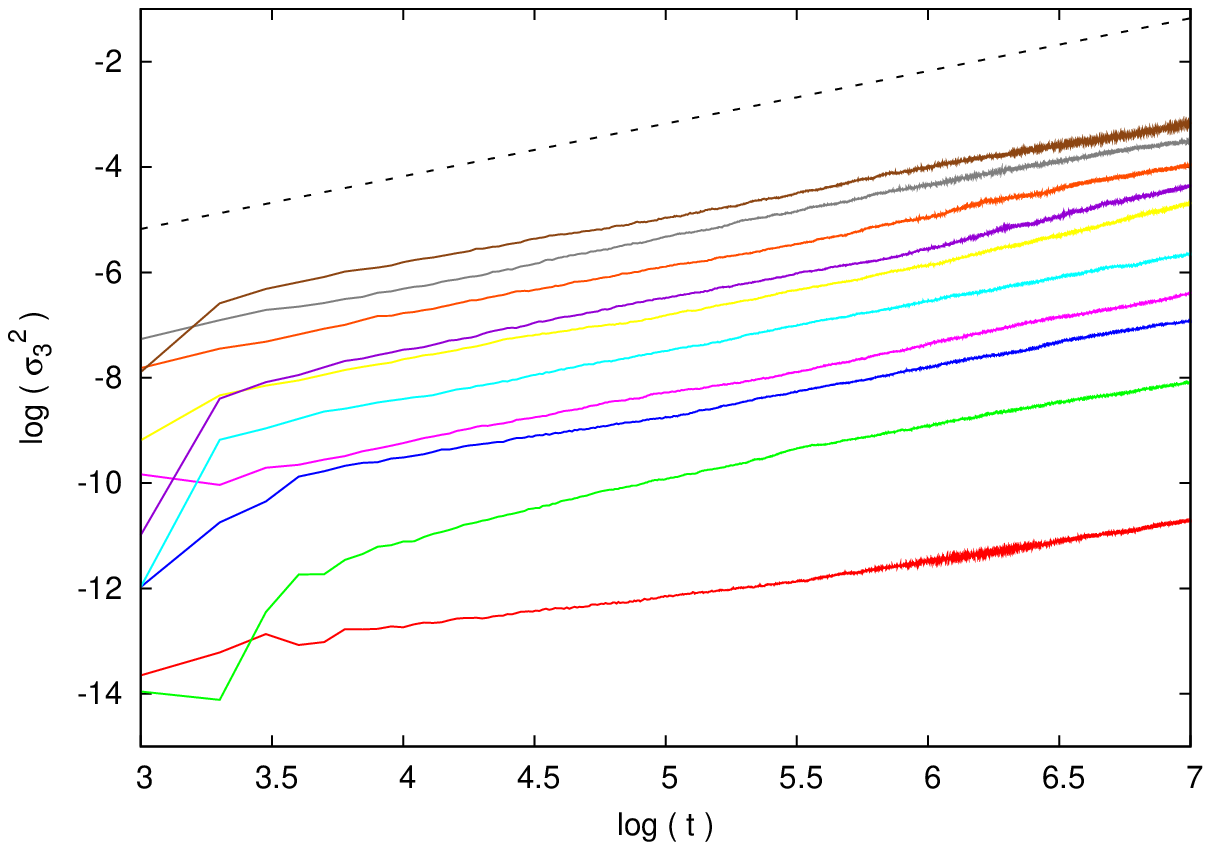}}
        \caption{Variance evolution for all the parameter values in logarithmic scale, for $t\in[10^3,10^7]$. The relation $\epsilon$--color
          is: $0.003$--red, $0.005$--green,  $0.007$--blue, $0.008$--magenta, $0.010$--cyan, $0.012$--yellow, $0.013$--violet,
          $0.015$--orange, $0.018$--grey, $0.003$--brown. The slopes have been fitted according to ansatz $\sigma_3^2(t)=c t^\eta$,
          for the interval $t\in[10^4,10^7]$, obtaining the values of $\eta$ given in Table~\ref{table: Hurst}.
          The dashed curve corresponds to a straight line of unitary slope plotted for comparison. 
          For most of $\epsilon$ values the diffusion is close to be normal.
        }%
      \label{fig: Variance_evolution_Log}
  \end{center}
\end{figure}
We have fitted power laws, according to the ansatz given at the end of 
Subsec.~\ref{subsec: statistical_quant}, for each element of $\mathcal{E}$, 
obtaining the values of the exponent $\eta$ shown in Table \ref{table: Hurst}. 
These values are not far from $\eta\approx 1$. Thus, from 
Figs.~\ref{fig: new_eps0.015},~\ref{fig: new_eps0.010}  
and~\ref{fig: new_eps0.005}, as well as the corresponding plots for the
rest of the values of $\epsilon$, that present a similar behavior, 
we conclude that the assumption of normal diffusion is a good first 
approximation. However, in some cases we find values appreciably smaller 
than 1, i.e., a slight sub-diffusive behavior. This we propose as a subject 
for future study. 

\begin{table}[h!]
\centering
\begin{tabular}{|c|c|c|c|c|c|c|c|c|c|c|}\hline
$\epsilon$ & $  0.003  $  & $  0.005  $ & $  0.007  $ & $  0.008$   & $  0.010  $ & $  0.012  $ & $  0.013  $ & $  0.015  $ & $  0.018  $ & $  0.020  $ \\%
$\eta$ & $0.75$& $0.87$& $0.91$& $0.95$& $0.91$& $1.10$& $1.18$& $0.98$& $0.89$& $0.86$ \\
\hline 
\end{tabular}
\caption{Power law behavior: values of exponent for the ansatz $\sigma_3^2(t)=c t^\eta$ 
for the interval $t\in[10^4,10^7]$.}
\label{table: Hurst}
\end{table}
We should note at this point that, although Chirikov's estimates of the theoretical 
diffusion coefficient concern time-averages as in (\ref{eq53a}), the assumption that 
time-averages and space-averages are similar is used along his analytical derivation 
of $D_C$. In fact, the aim of introducing the reduction factor $F$ to take into
account heuristically the correlations of the driving phases at the borders
of the stochastic layer, but assuming that the time-variance grows linearly with 
time.  However, a numerical derivation of $D$ is in practice only possible 
considering ensemble averages. In the lack of sufficient information about 
the correlations of the driving phases, we will then simply set $F=1$ in 
the comparison of $D$ with $D_C$ below. 

The fitted values of the numerical diffusion coefficients $D$, computed 
according to Eq.~(\ref{intro_dif_coef_variance}) using two different 
time intervals:  $[0,10^7]$ and $[10^6,10^7]$, are shown in the second and 
third columns of Table~\ref{table: D}, respectively. Both computations of 
$D$ yield quite similar values. In the fourth column we include the theoretical 
estimation $D_C$, given in table~\ref{table: D_C}, for comparison. 
\begin{table}[h!]
\centering
\begin{tabular}{cccc}
\hline\hline
\vspace*{-2ex} \\ %
$\epsilon$\hspace{4mm} &   $D~(0<t<10^7)$   &    $D~(10^6<t<10^7)$ & $D_C$ \\ %
\hline %
 $0.020$\hspace{4mm}   &   $6.7\times 10^{-11}$   &    $7.0\times 10^{-11}$   &  $8.4\times 10^{-11}$\\%
\hline  %
 $0.018$\hspace{4mm} &    $3.5\times 10^{-11}$    &    $3.5\times 10^{-11}$   &  $3.0\times10^{-11}$\\%
\hline  %
 $0.015$\hspace{4mm} &    $1.1\times 10^{-11}$    &    $1.1\times 10^{-11}$    &   $5.0\times10^{-12}$\\%
\hline  %
 $0.013$\hspace{4mm} &   $4.2\times 10^{-12}$     &    $4.2\times 10^{-12}$   &   $1.3\times10^{-12}$\\%
\hline  %
 $0.012$\hspace{4mm} &  $1.9\times 10^{-12}$      &    $1.9\times 10^{-12}$   &   $5.9\times10^{-13}$\\%
\hline  %
 $0.010$\hspace{4mm} &   $2.3\times 10^{-13}$     &    $2.3\times 10^{-13}$   &   $1.2\times10^{-13}$\\%
\hline  %
 $0.008$\hspace{4mm} & $4.0\times 10^{-14}$       &    $4.0\times 10^{-14}$   &   $1.9\times10^{-14}$\\%
\hline  %
 $0.007$\hspace{4mm} & $1.3\times 10^{-14}$       &    $1.3\times 10^{-14}$   &   $6.9\times10^{-15}$\\%
\hline  %
 $0.005$\hspace{4mm} &  $8.9\times 10^{-16}$      &    $8.9\times 10^{-16}$   &   $6.7\times10^{-16}$\\%
\hline  %
 $0.003$\hspace{4mm} &  $2.1\times 10^{-18}$        &  $2.1\times 10^{-18}$     &   $2.4\times10^{-17}$\\%
\hline\hline  \vspace*{-4ex} %
\end{tabular}
\caption{Diffusion coefficients for $\epsilon\in{\mathcal{E}}$. The second and third columns give
the values of the numerical coefficient, D, using the time intervals $[t_0,t_f]$ equal
to $[0,10^7]$ and $[10^6,10^7]$, respectively. In the fourth column we have added the theoretical
estimation, $D_C$, given in table~\ref{table: D_C}.} 
\label{table: D}
\end{table}
In Fig.~\ref{fig: comparaciones}--left we superpose all different estimates of the diffusion 
coefficient in semilogarithmic scale. The values of $D$ fitted with $0\leq t \leq 10^7$
and $D_C$, are displayed in colors red 
and green, respectively. Both the theoretical and the numerical coefficients have
nearly the same functional behavior with respect to $\epsilon$. Moreover, we notice 
that $0.1\lesssim D/D_C\lesssim 3.4$ for $\epsilon\in\mathcal{E}$,  which implies that 
the theoretical and numerical estimates agree rather well; the lower bound $0.1$ 
corresponding to $~\epsilon=0.003$ as can be seen from Figure~\ref{fig: comparaciones}--middle.  
The dots in this figure correspond to the values $D=D(D_C)$ in logarithmic scale, 
for $\epsilon\in\mathcal{E}$. Considering only parameter values: $0.005\leq 
\epsilon \leq 0.015$, we made a least square fit of the ansatz $\log(D)=q\log(D_C)+r$, 
obtaining an exponent $q=1.08333$. Thus, we conclude that the relation $D\sim D_C$ 
essentially holds true.
Notice that while performing the fit of the coefficients we explicitly discarded the values for
$\epsilon=0.003, 0.015, 0.020$. In fact, in the case of the lowest $\epsilon$ value, 
there are several terms in the perturbation of the very same order of magnitude, 
so that the layer resonance, that should have a coefficient $W_{\bm{l}}$ much
larger than the ones of the leading driving resonances, is actually not well
defined accordingly to Chirikov's theory (see Table~\ref{tabla1} and related discussion).
Meanwhile, for the largest $\epsilon$ values,
an overlap between the layer resonance and some very close high order resonances 
is observed, since the variance in $p_1$, $\sigma_1$, presents at large times, 
a slight increasing behavior, indicating that the width of the stochastic layer 
is not bounded, in contradiction with Chirikov's formulation. 
\begin{figure}[ht!]
      \fcolorbox{white}{white}{\hspace{-2.7mm}\includegraphics[width=0.37\textwidth]{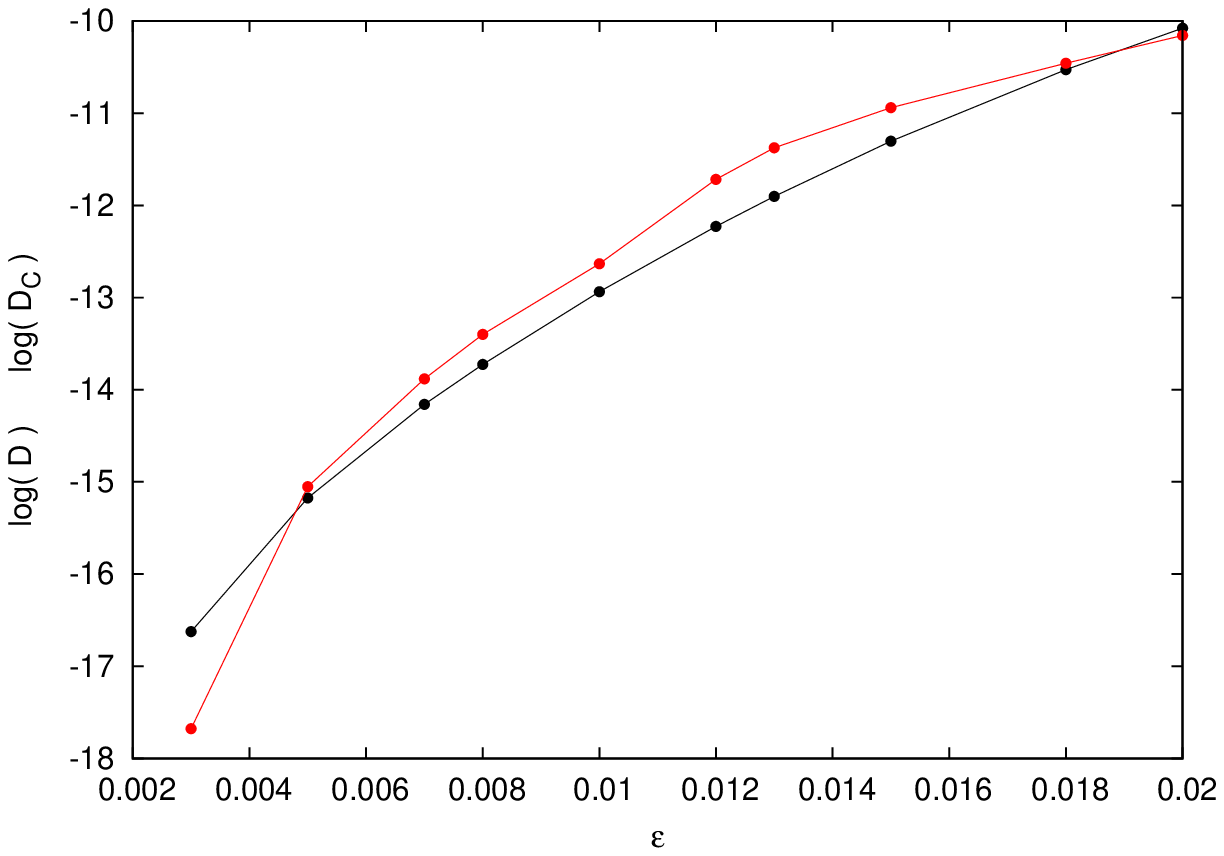}}
      \fcolorbox{white}{white}{\hspace{-12.5mm}\includegraphics[width=0.37\textwidth]{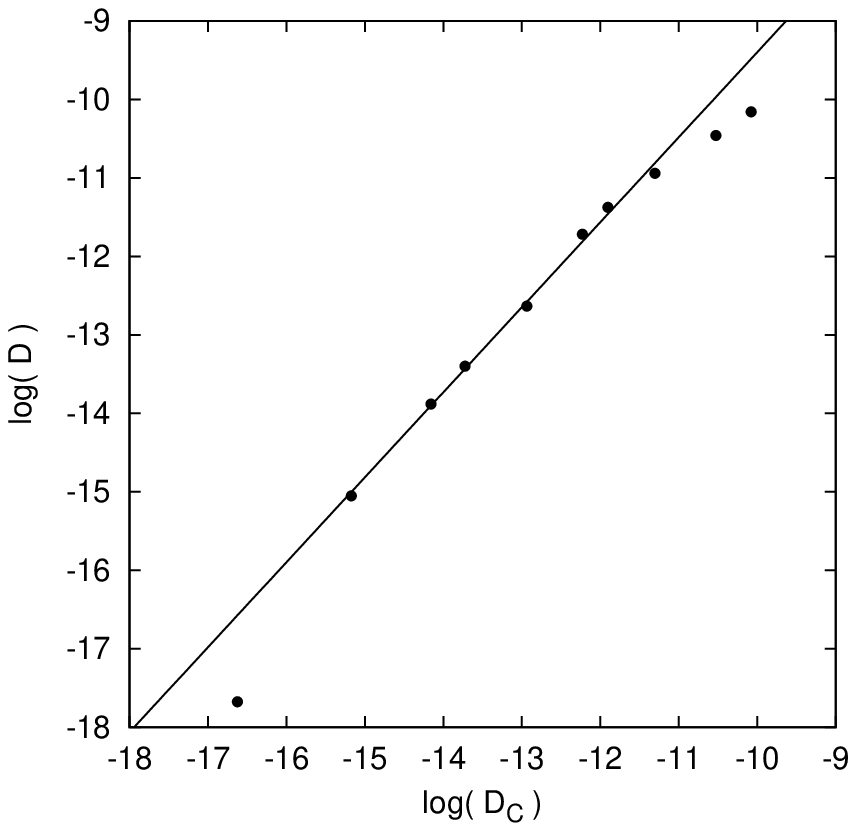}}
      \hspace{-2mm}\fcolorbox{white}{white}{\hspace{-9.5mm}\includegraphics[width=0.37\textwidth]{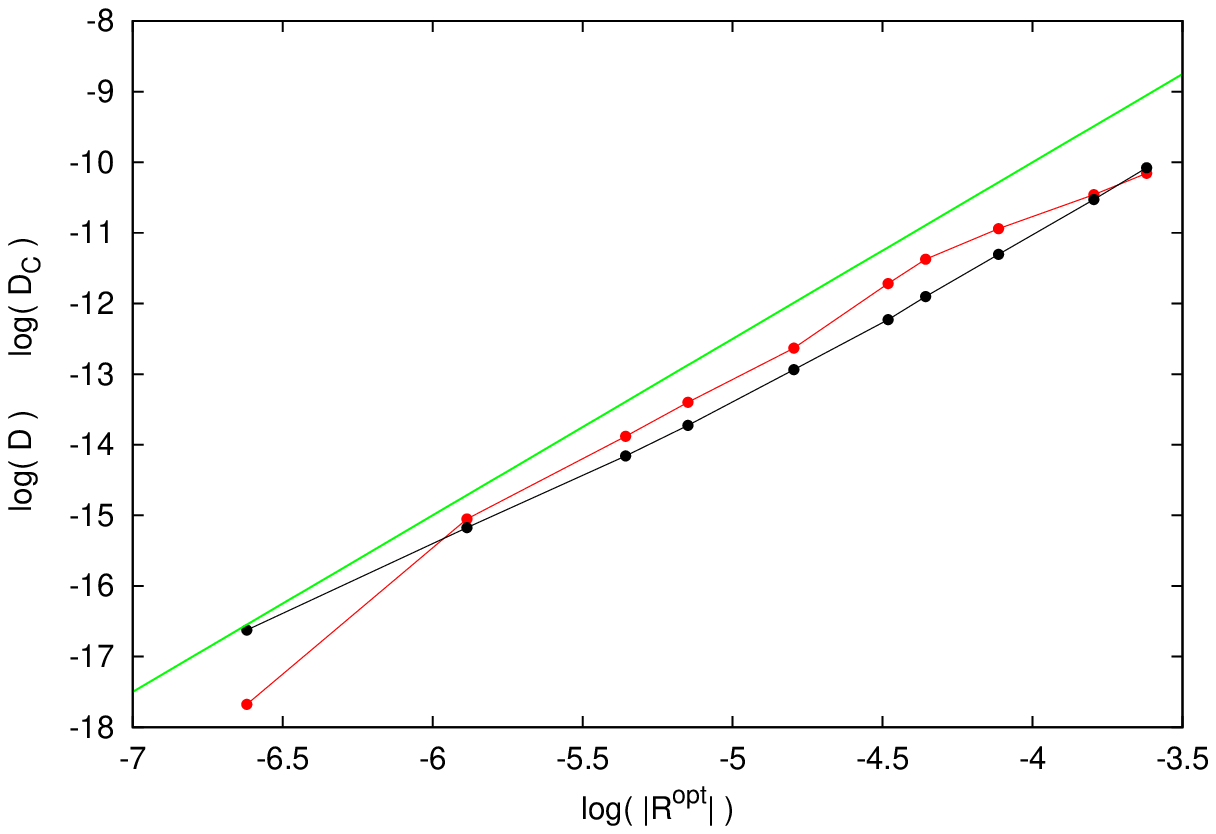}}
    \caption{{The left side panel shows the numerical and theoretical estimates of 
the diffusion coefficients 
      extracted from table~\ref{table: D}: $D$ ($0\leq t \leq 10^7$) in red  and $D_C$ in black.}
      The middle panel shows the comparison between the numerical and the theoretical diffusion
      coefficients. The line is the least square fit of the ansatz
      $\log(D)=q\log(D_C)+r$, with  $q=1.08333$. 
      The right side panel shows the diffusion coefficients versus the norm of the optimal remainder.
      The green line corresponds to the functions
      $||R_{opt}||^{2.5}$ and is displayed for comparison with the diffusion coefficients.
      It can be seen that both coefficients have an approximate functional
      trend of the shape $D,D_C\sim||R_{opt}||^{2.5}$, in black and red respectively.
    }
    \label{fig: comparaciones}
\end{figure}

As a final estimate, we have performed several fits of the ansatz $D,~D_C~\propto ||R^{opt}||^b$, 
considering different subsets of $\mathcal{E}$. Although the results depend on the chosen subset,
the value $b\approx2.5$ represents the mean situation. Fig.~\ref{fig: comparaciones}--right 
shows the corresponding behavior of  $D_C$ (red) and  $D$ (black). The green line corresponds 
to the function $||R_{opt}||^{2.5}$,  and it is displayed as a `guide to the eye' law 
for comparison with the laws found for various estimates of the diffusion coefficient. 

\section{Conclusions}\label{sec6}

In the present paper, we presented theoretical and numerical results pointing 
towards an important connection between the estimates for the diffusion rate 
along simple resonances in multidimensional nonlinear Hamiltonian systems that 
can be obtained using i) the theory of Chirikov ~\cite{Ch79}, and ii) the 
theory of Nekhoroshev ~\cite{N77}. We emphasized that, despite a common impression, 
the two theories are complementary rather than antagonist. In fact, we exploited 
this complementarity in order to obtain accurate theoretical predictions for the 
value of the diffusion coefficient along a resonance in a particular numerical 
example. Our main conclusions can be summarized as follows:

1) The theory of Chirikov {\it requires}, as a starting point, the construction 
of a simply-resonant normal form valid in local domains of the action space, 
which has to be optimal in the Nekhoroshev sense. In particular, the so-called 
driving terms of Chirikov's theory are identified with the remainder terms 
of the optimal normal form construction. Then, for small enough nonlinear 
perturbations $\epsilon$, the size of the driving terms turns to be of order 
$O(\exp(-1/\epsilon^a))$ (for some exponent $a>0$). 

2) We constructed the above optimal normal form in the so-called 3D quartic 
oscillator model \cite{CGS03}, using a computer-algebraic program in order 
to implement the normalization algorithm suggested in ~\cite{E08}. We were 
able to reach the optimal normalization order, at which the size of the 
normal form remainder becomes the least possible, and also to observe the 
expected asymptotic character of the normal form series. 

3) We used the computed expression for the optimal normalized Hamiltonian in 
order to transform all data in the basis of action variables suggested by Chirikov 
(see ~\cite{C02}). In this way, we identified the main resonant terms in the 
latter theory corresponding to i) the layer resonance, and ii) the driving 
resonances. With the above information at hand, we finally implemented 
Chirikov's formulae and computed a theoretical estimate $D_C$ for the 
diffusion coefficient along the resonance.

4) We compared the values of $D_C$ with a purely numerical measurement of the 
diffusion coefficient using ensembles of orbits integrated along the resonance's 
chaotic layer up to a quite long time. We found that for small perturbation 
values $\epsilon$, the diffusion can only be measured after subtracting, 
from the numerical orbital data, the so-called `deformation' effects. 
This requires transforming the data in new canonical variables arising 
by the same normalizing transformation that leads to the construction of 
the optimal normal form. 

5) We compared the theoretical prediction $D_C$ with the numerical value of 
the diffusion coefficient $D$ for various values of $\epsilon$. We found a 
quite satisfactory agreement, so that essentially one has $D_C\propto D$. 
The coefficient of proportionality depends on the so-called reduction factor 
of Chirikov's theory. In the lack of sufficient information about the value 
of the reduction factor, we simply set it equal to unity, which proves to 
be an adequate approximation. 

6) We pointed out that in the framework of Chirikov's theory one obtains a 
power-law relation between $D$ and the size of the optimal remainder 
$||R_{opt}||$, i.e. we have $D\sim ||R_{opt}||^{2+p}$, where $p$ is a 
positive constant of order unity. This is in agreement with the analysis 
made in \cite{EH12}. We applied this relation by a power-law fitting on 
our numerical data. We found that the heuristic law $D\sim ||R_{opt}||^{2.5}$ 
is adequately precise for all practical purposes. This latter fact allows 
to estimate directly the value of the diffusion coefficient using only 
normal form data. 

7) Finally, we made a preliminary study of the character of the diffusion 
along a resonance in the weakly chaotic regime. Although to a first 
approximation the diffusion can be characterized as normal, we found 
secondary features in all diffusion curves, that we attribute to the 
passage of some orbits of our ensemble from crossing points with 
secondary resonances. Furthermore, we tried power-law fitting of all 
the diffusion curves $\sigma^2\sim t^w$, where $\sigma^2$ is the variance 
of the ensemble in the action variables suggested by Chirikov. We found 
values of $w$ close to 1, but with a slight preference towards values 
smaller than unity. Thus, we conclude that the overall effect of secondary 
resonances is to render the chaotic spreading slightly sub-diffusive. 
This subject, however, necessitates a focused study that is proposed 
as a future subject.    

\section* {Acknowledgements}
PMC, CMG and MFM were supported with grants from the \emph{Consejo de
Investigaciones Cient\'{\i}ficas y T\'ecnicas de la Rep\'ublica Argentina},
and
the \emph{Universidad Nacional de La Plata}. CE acknowledges the financial
support and the hospitality of the \emph{Facultad de Ciencias Astron\'omicas
y Geof\'{\i}sicas-IALP}, where this research has been conducted. He has 
also been supported in part by the Research Committee of the Academy of 
Athens.  
The authors are very grateful to the two referees for the useful comments,
suggestions and criticism that serve to improve substantially the manuscript.


\appendix
\section{Chirikov's main derivations}\label{ap1}
We start from the expression for 
the Hamiltonian (\ref{hamppsi2}), given in section \ref{sec4} namely, 
\begin{eqnarray*}
H(\bm{p},\bm{\psi})&=&{p_1^2\over 2M_G} 
+|\bm{\omega}^r|p_2 + \sum_{l=1}^3 \sum_{k+ l> 2}^3 {p_kp_l\over 2M_{kl}}
+ \epsilon V_G\cos\psi_1
+ \epsilon \sum_{\bm{m}} \tilde{V}_{\bm{m}}
\cos(\bm{m}\cdot\bm{\theta}(\bm{\psi}))~~
\end{eqnarray*}
where the coefficients $\tilde{V}_{\bm{m}}$ have constant values.

In absence of perturbation ($\tilde{V}_{\bm{m}}=0$), the components $p_k, k=2,3$ are 
local integrals of motion, whose value is equal to zero if $\bm{I}^r$ is a point of 
the orbit. Then, the Hamiltonian reduces to:
\begin{equation}
H(\bm{p},\bm{\psi})\approx H_1(p_1,\psi_1)+\epsilon \tilde{V}(\bm{\psi})\label{eq26},
\end{equation}
where:
\begin{equation}
H_1={p_1^2\over 2M_G}+\epsilon V_G\cos\psi_1\label{eq27}
\end{equation}
is the pendulum Hamiltonian for the guiding resonance,  and
the perturbing
phases $\bm{\theta}$ in $\tilde{V}$ are written in terms of the new components 
$\psi_k$.

To transform the phase variables, we
take into account that the dot product is invariant under a change of
basis.  Recalling that $\psi_k=\sum_l\Upsilon_{kl}\theta_l$
then, if $\bm{\nu}$ denotes the vector $\bm{m}$ in the new basis, we have:
$\varphi_{\bm{m}}\equiv\bm{m}\cdot\bm{\theta}=\bm{\nu}\cdot\bm{\psi}$, where $\nu_k=
\sum_im_i\Upsilon_{ik}$. As we can readily see,
while the $m_k$ are integers, the quantities $\nu_k$
are, in general, real numbers.

As mentioned above, for $\tilde{V}=0$  the
$p_k$ are local integrals of motion and recalling that $H_1$ is also
an unperturbed integral, we have the full set of three local
integrals: $H_1, p_2, p_3$. But if we switch on
the perturbation, they will change with time.  This variation
is determined by the time dependence of $\varphi_{\bm{m}}$.
To get $\varphi_{\bm{m}}(t)$ we evaluate the dot product
$\bm{\nu}\cdot\bm{\psi}$:
\begin{equation}
\varphi_{\bm{m}}(t)=\bm{m}\cdot\bm{\theta}=\bm{\nu}\cdot\bm{\psi}\approx
\xi_{\bm{m}}\psi_1(t)+\omega_{\bm{m}}t+
\beta_{\bm{m}}, \label{eq2a9}
\end{equation}
where:
\begin{equation}
\xi_{\bm{m}}=\sum_{k=1}^3 {\nu_k(\bm{m})\over M_{k1}},\quad
\omega_{\bm{m}}\!=\!\bm{m}\!\cdot\!\bm{\omega}^r\!=\!\nu_2(\bm{m})|\bm{\omega}^r|;
\label{eq30a}
\end{equation}
and $\beta_{\bm{m}}$ is a constant. 

It can be found that the change in the unperturbed integrals over a 
half period of oscillation 
$T$, given by
\begin{equation}
T(w)={1\over\Omega_G}\ln(32/|w|),\qquad w=\frac{H_1}{\epsilon V_G}-1,
\label{eq43a}
\end{equation}
is
\begin{equation}
\Delta p_i\approx{\epsilon\over\Omega_G}\sum_{\bm{m}}{\nu}_i(\bm{m})
Q_{\bm{m}}\sin\varphi^0_{\bm{m}};\qquad i=2,3,
\label{eq44a}
\end{equation}
while
\begin{equation}
\Delta H_1\approx -{\epsilon|\bm{\omega}^r|\over\Omega_G}\sum_{\bm{m}}
{\nu_1(\bm{m})
\nu_2(\bm{m})\over\xi_{\bm{m}}}Q_{\bm{m}}\sin\varphi^0_{\bm{m}}.
\label{eq45a}
\end{equation}
The variation of the integrals depends on
$Q_{\bm{m}}=\tilde{V}_{\bm{m}}A_{2|\xi_{\bm{m}}|}(\lambda_{\bm{m}})$ where
$A_m(\lambda)$ denotes the  Melnikov--Arnold integral given by 
\begin{equation*}
A_m(\lambda)=\int_{-\infty}^{\infty} d\hat t
\cos\left({m\over 2}\psi^s (\hat t)-\lambda\hat t\right),\qquad
\psi^{s}(\hat t)=4\arctan \left(e^{{\hat t}}\right),
\end{equation*}
whose asymptotic value
 for large $\lambda$, is 
\begin{equation}
A_m(\lambda)\approx {4\pi(2\lambda)^{m-1}\over (m-1)!}
e^{-\pi\lambda/2}, \qquad\lambda\gg m.\label{eq14}
\end{equation}
The factorial should be replaced by the
Gamma function, $\Gamma (m)$, for non--integer $m$. For
details, see for instance ~\cite{FM07}, ~\cite{C02}.

Having already attained  the change in the integrals after a period of motion $T$, 
we need to compute the variation for the phases $\varphi^0_{\bm{m}}$ 
which are the quantities 
(\ref{eq2a9}) evaluated in the separatrix,  
 over the 
same time interval, in order to obtain a map describing
 Arnold diffusion. We have

\begin{eqnarray*}
\Delta\varphi^0_{\bm{m}}=\xi_{\bm{m}}\Delta\psi^s_1(t_0)+
\omega_{\bm{m}}\Delta t_0 = \omega_{\bm{m}}T(w)+C_{\bm{m}},
\end{eqnarray*}
where $C_{\bm{m}}$ is a constant.
Rewriting (\ref{eq45a}) in terms of
the dimensionless energy $w$ instead of $H_1$, we arrive at the following
map:

\begin{equation}
\bar w=w-{|\bm{\omega}^r|\over\Omega_G}\sum_{\bm{m}}W_{\bm{m}}
\sin\varphi^0_{\bm{m}},
\label{eq47a}
\end{equation}
\begin{equation}
\bar\varphi^0_{\bm{m}}=
\varphi^0_{\bm{m}}+\omega_{\bm{m}}T(\bar w)+C_{\bm{m}},
\label{eq48a}
\end{equation}
where

\begin{equation*}
W_{\bm{m}}=\frac{\nu_1(\bm{m})\nu_2(\bm{m})Q_{\bm{m}}}
{\xi_{\bm{m}}V_G} 
\end{equation*}
is the very same coefficient given by (\ref{wm}).
In the above map the bar indicates
the values
of the variables after crossing the surface $\psi_1=\pm\pi$.

The mapping given by (\ref{eq47a})--(\ref{eq48a}) is, in some sense,
similar to the
Whisker or Separatrix map, which has the following expression

\begin{equation}
\bar w=w+W\sin\tau_0, \qquad
\bar\tau_0=\tau_0 +\omega_{\bm{l}}T(\bar w)\quad\mathrm{mod}(2\pi),
\label{eq18a}
\end{equation}
where $W$ is a perturbation parameter like the $W_{\bm{m}}$'s,
$\omega_{\bm{l}}=\bm{m_l}\cdot\bm{\omega}$  and $\tau_0$ is the
phase of the perturbation.
This map describes the motion in the vicinity of the separatrix of the pendulum under
a perturbation. 

A well known result of this mapping is that the separatrix 
becomes a  chaotic layer of width, $w_s$. 
In other words, the change of the pendulum energy under a small
perturbation turns out to be  bounded. Experimentally, this bound, $w_s$, is due 
to the strong correlations of the phases $\tau_0$ for
large times (we refer to ~\cite{Ch79} for details).

Taking into account the similarities of the mappings (\ref{eq47a}) and (\ref{eq18a}), Chirikov
argues that the largest term in (\ref{eq47a}) leads to the so-called layer resonance. In fact
this conjecture seems to be true, considering the results given in Table~\ref{tabla1}.

Finally, the scalar diffusion coefficient along the direction of the vector  ${\bm{\mu}_3}$ 
is defined as
\begin{equation}
D={\overline{\Delta p_3(t)^2}\over T_a},
\label{eq53a}
\end{equation}
where $\overline{X(t)}$ denotes time average and 
$T_a$ is the mean period of motion within the
stochastic layer of the guiding resonance defined by
\begin{eqnarray}
T_a(w_s)\approx\int_0^{w_s} T(w)dw\approx {1\over\Omega_G}\ln(32e/w_s),
\nonumber
\end{eqnarray}
with $w_s\approx|\bm{\omega^r}|W_{\bm{l}}|\omega_{\bm{l}}|/\Omega_G^2\,\,$ and 
$T(w)$ given by (\ref{eq43a}) (see CH79 for details).
Therefore, from (\ref{eq44a}) for $i=3$ and (\ref{eq53a}) the formula
for Chirikov's diffusion coefficient given in (\ref{difchifull}) can be derived. 

\label{lastpage}
\end{document}